\documentclass[sigconf]{acmart}

\providecommand{\compilemode}{1}

\usepackage[utf8]{inputenc}   % to read UTF-8 text
\usepackage[T1]{fontenc}      % better font encoding
\usepackage{kotex}
\usepackage{graphicx}
\usepackage{subcaption}
\usepackage{float}  
\usepackage[export]{adjustbox}
\usepackage{multirow}
\usepackage{makecell}
\usepackage{tabularx}
\usepackage{booktabs}
\usepackage[skip=10pt]{caption}
\usepackage{tikz}
\usetikzlibrary{calc}
\usepackage{xcolor}
\usepackage{xspace}
\usepackage{listings}
\usepackage{enumitem}
\usepackage{algorithm}
\usepackage{algpseudocode}
\usepackage{amsmath}
\usepackage{amssymb}
\usepackage{stmaryrd} % \llbracket, \rrbracket used in §3 (TxRA semantics)
\usepackage[capitalise,noabbrev]{cleveref}
\usepackage{placeins}

\usepackage{multicol}
\usepackage{fancyhdr}

\definecolor{myred}{RGB}{200,0,0}
\usepackage{xcolor}

\AtBeginDocument{%
  }

\setcopyright{acmlicensed}
\copyrightyear{2018}
\acmYear{2018}
\acmDOI{XXXXXXX.XXXXXXX}

\acmConference[SIGMOD '27]{}{June 03--05,
  2018}{Woodstock, NY}

\acmISBN{978-1-4503-XXXX-X/2018/06}

\begin{document}

% Author-comment macros flattened (no color) for submission-ready rendering.
% To re-enable colored markup during drafting, restore \textcolor wrappers.
\newcommand{\wshan}[1]{#1}
\newcommand{\jhha}[1]{#1}
\newcommand{\yjpark}[1]{#1}
\newcommand{\revision}[1]{#1}
\newcommand{\OURSYSTEM}[1]{\texttt{CADENZA}\xspace}
\newcommand{\UPQUAL}[1]{0.40}
\newcommand{\UPCOST}[1]{37.2}
\newcommand{\UPLAT}[1]{11.2}
\newcommand{\para}[1]{\medskip\noindent\textbf{#1}}
\newcommand*\circled[1]{\tikz[baseline=(char.base)]{
    \node[shape=circle, fill=black, text=white, draw=black, inner sep=0.5pt] (char) {#1};}}
\newcommand{\TxRA}[1]{\texttt{TxRA}}

\newcommand{\map}[1]{\texttt{Map}}
\newcommand{\filter}[1]{\texttt{Filter}}
\newcommand{\join}[1]{\texttt{Join}}
\newcommand{\groupby}[1]{\texttt{GroupBy}}
\newcommand{\topk}[1]{\texttt{Top-K}}

\newcommand{\semfilter}[1]{\textsf{SemFilter}}

\newcommand{\semop}[1]{$SO$}
\newcommand{\groundtruth}[1]{$GT_{D_s}$}
\newcommand{\samples}[1]{$D_s$}
\newcommand{\samplesize}[1]{$|D_s|$}
\newcommand{\physicalplan}[1]{$P_P$}
\newcommand{\dataset}[1]{$D$}

\newcommand{\BIODEXMAXLAT}[1]{9.87}
\newcommand{\BIODEXMAXCOST}[1]{152}
\newcommand{\BIODEXMAXQUAL}[1]{0.06}

\newcommand{\TEXTEMBEDDINGSMALL}[1]{\texttt{Text-Embedding-3-Small}}
\newcommand{\GPTFOUR}[1]{\texttt{GPT-4.1}}
\newcommand{\GPTFOURMINI}[1]{\texttt{GPT-4.1-mini}}
\newcommand{\GPTFOURNANO}[1]{\texttt{GPT-4.1-nano}}
\newcommand{\GPTAUDIOMINI}[1]{\texttt{gpt-audio-mini}}

\ifnum\compilemode<2

% --- Round 2 revision cover letter (response to meta-reviewer + R1-R4) ---
% \input{revision/summary-of-revision}

% \clearpage

% --- Round 1 author response (kept for side-by-side comparison while drafting) ---
% \input{revision/author-feedback}

\title{CADENZA:  Compiling Natural-Language Intent into Task-Specific Operator DAGs for Semantic Query Processing}

\author{Jaehyun Ha}
\email{jhha@dblab.postech.ac.kr}
\affiliation{
  \institution{GSAI, POSTECH}
  \city{Pohang}
  \country{Korea}
}

\author{Yongjoo Park}
\email{yongjoo@illinois.edu}
\affiliation{
  \institution{Univ.~of Illinois Urbana-Champaign}
  \city{Urbana}
  \country{USA}
}

\author{Wook-Shin Han}
\authornote{Corresponding author.}
\email{wshan@dblab.postech.ac.kr}
\affiliation{
  \institution{GSAI, POSTECH}
  \city{Pohang}
  \country{Korea}
}

\begin{abstract}
Semantic query processing engines (SQPEs) extend relational query processing with \emph{semantic operators} that are executed via model inference over \revision{unstructured} data. Optimizing such queries is inherently multi-objective: model inference dominates latency and monetary cost, and outputs are stochastic and backend-dependent, so quality must be optimized alongside efficiency.
\revision{Existing SQPE optimizers do not expose each semantic operator instance's intermediate task outputs as a relational optimization object, leaving optimization unable to filter, reorder, route, threshold, or jointly tune them. We present \OURSYSTEM{}, which compiles each semantic operator \emph{instance}---a template bound to a natural-language \emph{intent}---into an intent-specific plan space of typed task DAGs and selects an executable plan under user-specified quality--latency--cost trade-offs.}
\OURSYSTEM{} introduces \emph{task-extended relational algebra} (\TxRA{}), a conservative extension of relational algebra with task-specific operators. \revision{The \emph{logical planner} synthesizes seed \TxRA{} plans, applies structural
rewrites whose safety conditions are checked from operator dependencies, and enumerates semantics-guided alternatives from alternative-generation templates.
The \emph{physical planner} compiles each task-specific operator into a router over heterogeneous backends and jointly tunes routing cutpoints, backend parameters, and relational thresholds with Bayesian optimization.}
On SemBench, \OURSYSTEM{} improves the scenario-level averages of quality, latency, and cost by up to +0.49, 165.7$\times$, and 310.3$\times$, respectively, relative to state-of-the-art.
\end{abstract}




\maketitle

\section{Introduction}  \label{sec:introduction}

With the advent of Large Language Models (LLMs), a new class of \emph{semantic query processing engines} (SQPEs) has emerged~\cite{LOTUS, abacus, AOP, unify, aws-redshift-ml, bigquery, palimpzest}. SQPEs extend relational query processing with \emph{semantic operators} whose behavior is specified in natural language and executed via model inference over \revision{unstructured} table data (e.g., text, images, audio). These operators mirror core relational operators (e.g., \texttt{SemFilter}, \texttt{SemJoin}, \texttt{SemMap}) while replacing hand-written UDF pipelines with prompt-driven matching, extraction, and reasoning as first-class operators. We distinguish a semantic operator template (a fixed algebraic operator such as \texttt{SemFilter}) from an instance, which binds the template to concrete inputs and an \emph{intent}---a natural-language specification of the operator's semantic goal. Figure~\ref{fig:intro:example} shows a representative query where \texttt{SemFilter} acts as a predicate over product images, flagging items whose package labels appear to contain compliance-sensitive claims.

\lstdefinelanguage{CustomSQL}{
  morekeywords={SemMap, SemJoin, SemFilter},
  sensitive=true,
  morestring=[b]',
}

\lstdefinestyle{sqldark}{
  language=CustomSQL,
  basicstyle=\ttfamily\small,
  numbers=none,
  numbersep=0pt,
  columns=fullflexible,
  keepspaces=true,
  showstringspaces=false,
  keywordstyle=\bfseries,      % only "model" will be bold
  stringstyle=\color{teal},    % strings in teal
  commentstyle=\itshape,
  rulecolor=\color{black},
  xleftmargin=10pt,
  framexleftmargin=0em,
  frame=none,
  upquote=true
}

\begin{figure}[t]
  \centering
\begin{lstlisting}[style=sqldark,label={fig:sql-semfilter}]
SELECT b.BrandName, COUNT(*) AS NumFlagged
FROM Products p JOIN Brands b ON p.BrandId = b.BrandId
WHERE p.Category IN ('Supplements','Cosmetics')
  AND SemFilter('Compliance-sensitive claims (e.g., "FDA 
  approved") appear on the package labels in {p.Image}.')
GROUP BY b.BrandName;
\end{lstlisting}
\vspace{-3mm}
\caption{A semantic operator instance embedded in SQL.}
\vspace{-8mm}
\label{fig:intro:example}
\end{figure}

Because semantic operators invoke expensive and inherently uncertain inference, SQPEs face a multi-objective optimization problem~\cite{abacus}. Model-backed steps can dominate end-to-end latency and monetary cost~\cite{LOTUS, abacus, thalamusdb}, while semantic outputs are non-deterministic and backend-dependent. Semantic query optimization is thus best viewed as selecting plans that balance quality, latency, and cost rather than minimizing a single metric.

\revision{Existing SQPE optimizers~\cite{palimpzest, abacus, LOTUS, AOP, unify, bigquery, DOCETL} do not expose each semantic operator instance's internal task structure as a relational optimization 
object, leaving optimization unable to filter, reorder, route across heterogeneous backends, threshold, or jointly tune its intermediate task outputs. Figure~\ref{fig:overview} illustrates four such patterns. \emph{Direct LLM} is a monolithic LLM call per row, the simplest supported execution pattern. \emph{Proxy Cascade} represents LOTUS-style~\cite{LOTUS} proxy execution, where a cheaper model or embedding score decides which tuples escalate to the LLM. 
\emph{Reduced Context} represents 
Palimpzest/Abacus-style~\cite{palimpzest, abacus} reduced-context execution, where the input is chunked and only the top-$k$ chunks 
are fed to the LLM. 
\emph{Pipeline Rewrite} represents DocETL-style~\cite{DOCETL} pipeline-level decomposition. None of these expose intermediate task outputs 
to relational optimization.}

\revision{Our key insight is that an intent-specific decomposition of a semantic operator instance into a DAG of typed sub-operators turns its intermediate task outputs into relational attributes that optimization can filter, reorder, route across heterogeneous backends, threshold, and jointly tune. For example, the filter in Figure~\ref{fig:intro:example} can be achieved by an OCR step that extracts text from each label, followed by a TxtQA step that checks whether the text contains a compliance-sensitive claim, where each step is served by a model whose capacity matches its difficulty.}

\revision{However, achieving this goal raises three challenges (C), each motivating a desideratum (D). 
(C1) Intent is given in open-ended natural language with no algebraic form, so the optimizer must (D1) synthesize an intent-specific DAG of typed sub-operators. (C2) The right backend for a sub-operator is input-dependent: some tuples are best handled by deterministic rules, others by specialized models, and the residual hard cases require LLMs---a heterogeneity that requires (D2) per-tuple dispatch over heterogeneous backends. (C3) The plan space is multi-objective and plan evaluation is expensive, so (D3) search must be sample-efficient and budget-bounded.}

We present \OURSYSTEM{}\footnote{In music, a cadenza is a structured yet improvisational passage that adapts the performance to the specific piece; analogously, \OURSYSTEM{} synthesizes and tunes a bespoke execution plan for each intent instance.}, a semantic operator optimizer that \revision{addresses these three desiderata.} \revision{Given an instance with its intent and a user-specified preference $(w_q, w_l, w_c)$ over quality, latency, and cost,} the \textit{logical planner} synthesizes \revision{an intent-dependent \emph{plan space} of sub-operator DAGs (D1)}, and the \textit{physical planner} compiles each sub-operator into tailored implementations \revision{(D2) and tunes via Bayesian optimization (D3)}. 
\jhha{\OURSYSTEM{} focuses on optimizing individual semantic operator instances; joint optimization across operators is orthogonal to our focus. Crucially, per-operator optimization already captures most gains: even on multi-operator queries, composing locally optimized instances substantially outperforms state-of-the-art baselines (\S\ref{sec:endtoendperf}).}

We formalize such sub-operators in \emph{task-extended relational algebra} (\TxRA{}), a conservative extension of relational algebra. \TxRA{} keeps \emph{\revision{relational}} operators (e.g., $\sigma,\pi,\bowtie$) intact for deterministic schema and dataflow manipulation, but adds \emph{task-specific} operators perform inference over typed attributes and expose their outputs as relational attributes \revision{(e.g., \textsf{OCR} returns text, and \textsf{TxtQA}/\textsf{VQA} return an answer with a confidence score, each appended to its input row)}.

The logical planner translates a semantic operator instance into a \TxRA{} DAG (i.e., a logical plan). Unlike traditional DBMSs, where the query explicitly determines the logical plan structure (e.g., SQL), its intent is given as open-ended natural language with no algebraic form.
\revision{To address D1,} the planner thus (i) synthesizes a small set of seed plans via an iterative draft--refine loop under a fixed LLM budget, and \revision{(ii) expands them with two mechanisms: \emph{correctness-preserving structural rewrites} and \emph{semantics-guided alternative generation}. Structural rewrites are dependency-based algebraic rewrites whose preconditions are checked from schemas and dataflow. In contrast, semantic alternatives such as decomposition and proxy substitution are not assumed to be equivalence-preserving; they enlarge the candidate plan space and are selected empirically using the validation/teacher signals used by the physical optimizer.} \revision{For example, Figure~\ref{fig:intro:example}'s query yields a seed plan such as $\text{Products} \to \textsf{VQA}(\text{Image}, q) \to \sigma_{\text{Answer}=\text{Yes} \wedge \text{Score}\geq\lambda}$, for which decomposition produces an alternative $\text{Products} \to \textsf{OCR}(\text{Image}) \to \textsf{TxtQA}(\text{txt}, q) \to \sigma_{\text{Answer}=\text{Yes} \wedge \text{Score}\geq\lambda}$, where $q$ is the predicate.} Crucially, LLM usage is confined to budgeted seed-plan generation; most exploration comes from the mechanisms.

The physical planner takes the candidate logical plans produced by the logical planner, finds the best physical plan for each logical plan, and selects the overall winner under the user’s quality, latency, and cost preferences. \revision{To address D2, the planner synthesizes a \emph{data-aware router} per task-specific operator: per tuple, it scores input difficulty and dispatches the tuple to one of heterogeneous backend families---\emph{symbolic} (deterministic rules), \emph{specialized} (compact task-tuned models), \emph{general-purpose} (foundation models), and \emph{composite} (multi-stage pipelines)---via tunable cutpoints. For instance, an \textsf{OCR} router may dispatch easy labels to a distilled OCR model and escalate difficult images to a vision LLM. To address D3, the planner jointly tunes the physical plan’s parameter space $\Theta$---consisting of the router’s cutpoints, per-backend knobs, and parameters of relational operators (e.g., score thresholds $\lambda$)---with Bayesian optimization under a fixed evaluation budget. \revision{While the data-aware router is not an isolated model-selection technique, our contribution is its placement and joint tuning: routing 
is attached to each task-specific operator and its cutpoints are tuned together with backend and relational parameters as one SQPE physical optimization problem.}} For deployment, we select a single Pareto-optimal plan by maximizing a user-weighted utility; to approximate the Pareto frontier, we instead optimize hypervolume. \revision{To ensure practicality, the planner also enforces runtime safety and manages optimization overhead.}

\revision{\OURSYSTEM{} does not cost every subplan as classical optimizers do: with no reliable closed-form cost/quality model for task-specific operators, costing a (sub)plan is not a cheap formula lookup but an execution on samples. Since each candidate is a short task-specific DAG, whole-plan BO evaluation is affordable, whereas subplan-level costing would require BO-style evaluation for many operator/sub-DAG fragments. Moreover, unlike dynamic-programming (DP)-style optimization or interleaved search that prunes plans via partial evaluation, our staged design is necessary because synthesized plans are open-ended and can violate DP's optimal-substructure assumption: intermediate outputs (e.g., noisy OCR) may be corrected downstream, so early pruning may discard the eventual winner.}

\OURSYSTEM{} is explicitly \emph{budget-aware} and \emph{anytime}: it
searches over candidate logical and physical plans under fixed evaluation budgets, returning the best plan found so far and improving monotonically with additional search. 
This is a deliberate design choice aligned with real deployments, where each additional candidate incurs model inference and evaluation cost. To control 
backend stochasticity without inflating evaluation cost, we run a single evaluation per BO trial and treat it as a noisy observation in the surrogate, then re-evaluate only the final top-$K$ physical plans for robust selection.

\para{Contributions.} This work makes the following contributions:
\begin{itemize} [leftmargin=4mm]
\item We present \OURSYSTEM{}, \revision{an SQPE-level optimizer that compiles a semantic operator instance into \TxRA{} plans and selects an executable plan under quality--latency--cost preferences}.
\item We introduce task-extended relational algebra, which adds task-specific operators to \revision{relational} operators, enabling semantic processing to be expressed and optimized as an operator DAG.
\item We develop a logical planner that \revision{synthesizes seed \TxRA{} plans, applies structural rewrites, and enumerates semantics-guided alternatives from alternative-generation templates.}
\item We propose a physical planner that compiles logical plans into physical plans with a data-aware router over family-specific implementations and tunes them with Bayesian optimization.
\item We evaluate \OURSYSTEM{} on SemBench~\cite{sembench} and show up to +0.49 quality and 165.7$\times$/310.3$\times$ latency/cost improvements in scenario-level averages over state-of-the-art baselines.
\end{itemize}

\vspace{1mm}
\noindent
Section~\ref{sec:overview} provides an overview of \OURSYSTEM{} and formalizes the optimization problem; Sections~\ref{sec:logical_planner}--\ref{sec:physical_planner} present the logical and physical planners; Sections~\ref{sec:implementation}--\ref{sec:experiments} describe implementation and evaluation; Section~\ref{sec:related_work} discusses related work; and Section~\ref{sec:conclusion} concludes.
\begin{figure*}[t]
\includegraphics[width=\linewidth]
{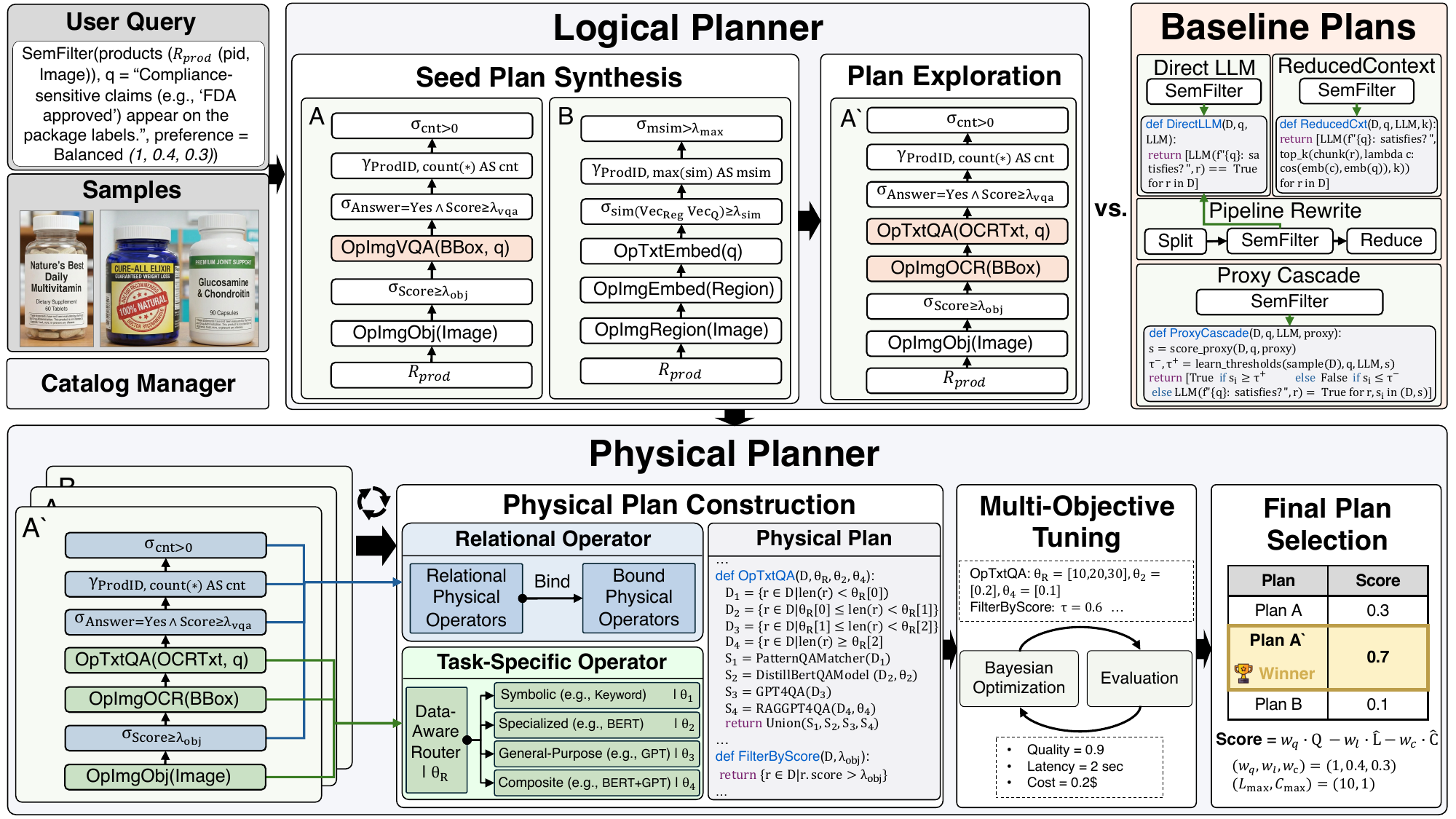}
\vspace{-6mm}
\caption{\revision{Overview of the \OURSYSTEM{} architecture illustrating an end-to-end workflow example with representative baseline plans. For clarity, implementation-level optimizations that apply uniformly across plans (e.g., LLM batching) are not shown.}}
\vspace{-4mm}
\label{fig:overview}
\end{figure*}

\section{An Overview of \OURSYSTEM{}} \label{sec:overview}

This section overviews \OURSYSTEM{}. We first formalize the semantic operator optimization problem and objectives (\S\ref{sec:overview:moop}), then present the system architecture through a motivating example (\S\ref{sec:overview:arch})).

\subsection{Semantic Operator Optimization Problem} \label{sec:overview:moop}

We formalize the optimization problem as finding a Pareto-optimal plan over quality, cost, and latency. For deployment, we select one Pareto-optimal plan which maximizes a user-weighted utility $U$.

\begin{itemize} [leftmargin=4mm]
    \item \textbf{Semantic Operator Instance ($\mathsf{SO}_{\mathcal{I}}$):}
    An instantiation of a semantic operator template with a natural-language \emph{intent} $\mathcal{I}$ that specifies the operator's semantic goal.
    It denotes an intended transformation $\mathsf{SO}_{\mathcal{I}} : \mathsf{Rel}_{\text{in}} \rightarrow \mathsf{Rel}_{\text{out}}$.

    \item \textbf{Logical Plan ($\mathcal{L}$):}
    A DAG of typed logical operators specifying \emph{what} computation performs $\mathsf{SO}_{\mathcal{I}}$.
    We denote the set of valid logical plans as $\mathbb{L}(\mathsf{SO}_{\mathcal{I}})$.

    \item \textbf{Physical Plan ($\mathcal{P}$):}
    A concrete, executable realization of a logical plan, where each operator is bound to a specific implementation and parameters (some tunable). The search space is the union of all such realizations:
    \[
    \mathbb{S}(\mathsf{SO}_{\mathcal{I}}) \;=\; \bigcup_{\mathcal{L}\in \mathbb{L}(\mathsf{SO}_{\mathcal{I}})}\mathbb{P}(\mathcal{L}).
    \]
\end{itemize}

\para{Optimization Goal.}
Given a semantic operator instance, we optimize over the plan space $\mathbb{S}$ under three competing objectives: maximize quality $Q(\mathcal{P})$ while minimizing latency $L(\mathcal{P})$ and monetary cost $C(\mathcal{P})$.
For deployment, we select a single plan by maximizing a user-weighted scalar utility
$U(\mathcal{P}) = w_q\,Q(\mathcal{P}) - w_l\,\hat{L}(\mathcal{P}) - w_c\,\hat{C}(\mathcal{P})$
for weights \revision{$w_q, w_l, w_c \ge 0$ (at least one positive; setting a weight to $0$ omits that metric)}, where $\hat{L}(\mathcal{P})$ and $\hat{C}(\mathcal{P})$ denote normalized latency and cost, respectively.
With \revision{all} positive weights, the selected plan is Pareto-optimal in the $(Q,L,C)$ trade-off space.

\para{Pareto Optimality of Utility Maximization.}
Suppose $\mathcal{P}^\star$ is not Pareto-optimal. Then there exists $\mathcal{P}'\in\mathbb{S}$ such that
$Q(\mathcal{P}') \ge Q(\mathcal{P}^\star)$, $\hat{L}(\mathcal{P}') \le \hat{L}(\mathcal{P}^\star)$, and $\hat{C}(\mathcal{P}') \le \hat{C}(\mathcal{P}^\star)$,
with at least one inequality strict.
Because $w_q,w_l,w_c>0$, we have
$U(\mathcal{P}') - U(\mathcal{P}^\star)
= w_q\!\big(Q(\mathcal{P}')-Q(\mathcal{P}^\star)\big)
+ w_l\!\big(\hat{L}(\mathcal{P}^\star)-\hat{L}(\mathcal{P}')\big)
+ w_c\!\big(\hat{C}(\mathcal{P}^\star)-\hat{C}(\mathcal{P}')\big) > 0$,
which contradicts the optimality of $\mathcal{P}^\star$. \qed

\para{\revision{Constrained Optimization.}}
\revision{Constrained-optimization formulations (e.g., $\max Q$ subject 
to $L \leq L_\mathrm{max}$) are also supported by modifying the BO procedure, e.g., by adding budget-violation penalties to the BO scoring objective (Appendix~\ref{sec:appx:constrained}); the 
utility-based interface is our default for its simplicity.}

\subsection{System Architecture} \label{sec:overview:arch}

% [CHANGE] Make inputs/outputs explicit (relations, catalog, prefs) + clarify the output artifact.
\OURSYSTEM{} takes as input (i) a semantic operator instance,
(ii) the input relation(s) containing \revision{unstructured} attributes, (iii) an optional validation set, and (iv) an optional user-specified preference over quality, latency, and cost.
It outputs an executable physical operator DAG---a fully bound plan with concrete implementations and tuned parameters.
As illustrated in Figure~\ref{fig:overview} (light gray boxes), the architecture consists of three main components:

\begin{description}  [leftmargin=0mm]
    \item[Catalog Manager] Maintains the operator catalog, which enumerates \revision{relational} and task-specific operators and implementations.

    \item[Logical Planner (\S\ref{sec:logical_planner})]
    Determines \emph{what} to compute by decomposing the user's intent into a DAG of typed logical operators.

    % [CHANGE] Keep BO mention lightweight here to avoid duplication with tuning paragraph below.
    \item[Physical Planner (\S\ref{sec:physical_planner})]
    Determines \emph{how} to compute it by compiling the logical plan into a physical plan with tunable parameters and tunes them via Bayesian Optimization.
\end{description}

% [CHANGE] Re-introduce catalog + validation set (missing in prior revision) to ground how tuning is done.

\para{\revision{User-Specified Preference.}}
\revision{The preference is implemented as weights $(w_q, w_l, w_c) \ge 0$ over quality, latency, and cost (a $0$ weight omits that metric). This weighted-utility formulation is a pragmatic deployment interface that provides a simple soft-preference mechanism for top-$k$ plan selection, consistent with prior work on rank-aware query optimization~\cite{rankawareqo, ranksql}. Fine-grained weights offer maximum flexibility, but for non-expert users we provide a discrete preset that maps to predefined weights---\emph{quality-first} $(1, 0.1, 0.1)$, \emph{balanced} $(1, 0.4, 0.3)$ (default), and \emph{efficiency-first} $(0.1, 1, 1)$.}

\para{Validation Set.}
Following prior work~\cite{LOTUS, abacus}, we estimate a plan’s quality, latency, and cost on a small labeled validation set.
If not provided, we sample tuples and label them once with a configurable teacher model; this one-time cost is included in optimization overhead and the labels are reused across all candidates.
\jhha{Our goal is not to predict exact accuracy but to rank candidate plans correctly. Empirically, this ranking remains stable even with moderate teacher noise (e.g., 10--20\% errors on hard samples), since bad plans (e.g., overly aggressive symbolic filters) fail on easy cases the teacher gets right, while good plans correlate well with the teacher’s decisions (See~\S\ref{sec:val_noise}).}
Final evaluations are performed separately.

\para{Logical Plan Exploration (Figure~\ref{fig:overview}, Top).}
The logical planner explores different strategies to realize the intent.
First, it generates a small set of initial seed plans.
\revision{under a 
fixed LLM budget. It then expands these seeds using two mechanisms: 
dependency-checked structural rewrites and semantics-guided alternative generation.}
In our running example:
\begin{itemize}  [leftmargin=4mm]
    \item \revision{\textbf{Plan A (direct).} Find candidate objects and
    ask VQA the intent \emph{directly} per object---``does this object
    satisfy $q$?''; a product passes if any object affirms it
    ($\gamma_{\texttt{ProdID},\,\mathrm{count}(*)} \rightarrow
    \sigma_{\texttt{cnt}>0}$).}
    \item \revision{\textbf{Plan B (proxy).} Embed image regions and the intent text
    into a shared space and ask a \emph{proxy} question---``is any region
    close to the intent?''; a product passes if its maximum region--intent
    similarity exceeds $\lambda_{\max}$
    ($\gamma_{\texttt{ProdID},\,\max(\mathrm{sim})} \rightarrow
    \sigma_{\texttt{msim}>\lambda_{\max}}$).}
\end{itemize}
\revision{Subsequently, \textbf{Plan Exploration} proposes \textbf{Plan 
A$'$}, which replaces the monolithic VQA operator with an $\text{OCR} 
\rightarrow \text{TxtQA}$ alternative (highlighted in red in 
Figure~\ref{fig:overview}). This alternative is not assumed to be 
equivalent a priori; it is compiled, tuned, and ranked under the same 
quality--latency--cost objective.}

\para{Physical Plan Construction (Figure~\ref{fig:overview}, Bottom).}
For each discovered logical plan, the physical planner compiles it into a physical plan with tunable parameters by binding each logical operator to a concrete physical implementation.
\begin{itemize}  [leftmargin=4mm]
    \item \textbf{\revision{Relational} Operators:}
    Predicates such as $\sigma_{\text{Score}\ge \lambda}$ are compiled into standard filters with a tunable parameter $\lambda$.

    \item \textbf{Task-Specific Operators:}
    Complex operators are compiled into an implementation that incorporates a data-aware router to dispatch tuples to family-specific implementations.
    For example, \textsf{OpTxtQA} in Plan A$'$ dispatches tuples across various implementations (symbolic matching, a distilled QA model, a strong general-purpose LLM, and a composite RAG-style),
    using cheap input features such as text length with a tunable threshold $\theta_{R}$, together with per-family parameters such as a confidence threshold ($\theta_2$) and a similarity threshold ($\theta_4$) (see ``Physical Plan'' in Figure~\ref{fig:overview}).
    When available, the compiler reuses the cataloged implementation for each family; otherwise, it generates the missing family-specific implementations and synthesizes the router.
\end{itemize}

\para{Multi-Objective Tuning.}
For each candidate logical plan, \OURSYSTEM{} tunes the parameters of its compiled physical plan using Bayesian optimization on the validation set.
The optimizer maximizes a scalar utility that trades off Quality ($Q$), Latency ($L$), and Cost ($C$).
We estimate $L_{\max}$ and $C_{\max}$ via an LLM-only execution on the validation set. For stability, we use logarithmic scaling (details in \S\ref{sec:physical_planner}).

\para{Final Plan Selection.}
After tuning each candidate logical plan, \OURSYSTEM{} compares the resulting best tuned physical plans and selects the overall winner by the same scalar utility.
As shown in the ``Final Plan Selection'' table, plan A$'$ is selected because its tuned physical plan preserves high accuracy via the OCR--TxtQA pipeline
while reducing cost relative to the VQA approach (Plan A).

\para{Stability under Stochastic Planning.} 
Model inferences can be stochastic, so repeated optimization runs may follow different trajectories and select a different final best plan for the same query.
We therefore focus on stability at the level of outcome metrics $(Q,L,C)$ (or a scalar utility) rather than exact plan identity.
To reduce run-to-run variance, we (i) control avoidable randomness using deterministic decoding when available, (ii) treat each BO trial as a noisy observation and (iii) smooth residual noise by re-evaluating the final top-$K$ candidates multiple times and selecting via an aggregated score (e.g., median/mean), optionally requiring an improvement margin $\epsilon$ before replacing the incumbent.

\section{Task-Extended Relational Algebra} \label{sec:algebra}

Semantic operator instances are specified by open-ended natural-language intents and executed via costly, uncertain inference over \revision{unstructured} data.
Treating an instance as a black-box UDF hides intermediate signals and blocks classical transformations such as pushdown and commutation.
Meanwhile, classical relational algebra cannot express semantics-dependent predicates over such modalities.
We therefore introduce \emph{task-extended relational algebra} (\TxRA{}), a conservative extension that encapsulates inference behind \emph{typed tasks} and exposes their outputs as relational attributes, enabling semantic processing to be optimized as an operator DAG.

\revision{As in relational algebra (RA), \TxRA{} separates the logical specification of an operator from its physical realization: a task-specific operator declares \emph{what task} to perform (e.g., \textsf{OCR}, extracting text from an image, with output schema $(\texttt{txt}{:}\textsf{Text})$), independent of \emph{how} it is implemented (e.g., for \textsf{OCR}, a vision LLM vs.\ Tesseract vs.\ a distilled OCR model). A \TxRA{} query thus forms an operator DAG amenable to two kinds of plan
exploration. \emph{Structural} rewrites, derived from schema and dependency
contracts, follow classical RA transformations such as pushdown and
commutation. \emph{Semantic} alternatives, such as $\textsf{VQA}$ to $\textsf{OCR}
\rightarrow \textsf{TextQA}$, are not assumed to be equivalence-preserving by
default. \S\ref{sec:exploration} details both mechanisms.}

\para{Types, Domains, and Carrier Set.}
Let $\mathbb{T}$ be a type universe, where each $\tau \in \mathbb{T}$ has a carrier domain $\mathcal{D}_\tau$.
Let $\mathbb{D}=\{\mathcal{D}_\tau \mid \tau \in \mathbb{T}\}$ be the set of such domains, partitioned as $\mathbb{D} = \mathbb{D}_{\mathrm{syn}} \;\cup\; \mathbb{D}_{\mathrm{sem}}$
where $\mathbb{D}_{\mathrm{syn}}$ contains structured domains supporting deterministic comparisons (e.g., \textsf{Int}, \textsf{Float}), and $\mathbb{D}_{\mathrm{sem}}$ contains domains of semantic objects (e.g., \textsf{Text}, \textsf{Image}).
We write $\mathsf{Rel}_{\mathbb{T}}$ for the set of typed relations whose attributes have types in $\mathbb{T}$ (equivalently, range over domains in $\mathbb{D}$). \revision{For example, in a product relation $r(\texttt{pid{:}Int},\ \texttt{img{:}Image})$, \texttt{pid} is in $\mathbb{D}_{\mathrm{syn}}$ and \texttt{img} is in $\mathbb{D}_{\mathrm{sem}}$.} Values in $\mathbb{D}_{\mathrm{sem}}$ support identity equality and syntactic predicates \revision{(e.g., on file size or hash for \texttt{img})}, while latent semantics-dependent predicates \revision{(e.g., on its label text or depicted brand)} are expressed via task-specific operators.

\para{Algebra Definition.}
We define $\TxRA{} = (\mathsf{Rel}_{\mathbb{T}},\ \Omega_{\mathrm{rel}} \cup \Omega_{\mathrm{task}})$, where $\Omega_{\mathrm{rel}}$ denotes \revision{relational} operators and $\Omega_{\mathrm{task}}$ denotes task-specific operators. All operators are closed over $\mathsf{Rel}_{\mathbb{T}}$.

\para{\revision{Relational} Operators ($\Omega_{\mathrm{rel}}$).}
$\Omega_{\mathrm{rel}}$ comprises standard relational operators (e.g., $\sigma,\pi,\bowtie,\gamma$) with well-typed deterministic predicates over $\mathbb{D}_{\mathrm{syn}}$.
For $\mathbb{D}_{\mathrm{sem}}$, we permit only deterministic predicates over raw content or metadata (e.g., exact match/regex/length for \textsf{Text}, and IDs/hashes/basic metadata for \textsf{Image}); thus, joins are keyed only by $\mathbb{D}_{\mathrm{syn}}$ attributes or deterministic identifiers from $\mathbb{D}_{\mathrm{sem}}$, while meaning-based matching is expressed via $\Omega_{\mathrm{task}}$.
Schema and cardinality semantics follow classical relational algebra.

\para{Task-Specific Operators ($\Omega_{\mathrm{task}}$).}
Let $\mathcal{T}$ be a task space. A \emph{task} $t \in \mathcal{T}$ is a typed (possibly stochastic) computation with signature
\[
\mathrm{sig}(t):(\tau_1,\ldots,\tau_m)\rightarrow \mathsf{Rel}_{\mathbb{T}},
\]
where inputs may include semantic types (e.g., \textsf{Text}, \textsf{Image}). \revision{For example, $\textsf{OCR}: \textsf{Image} \to (\texttt{txt}{:}\textsf{Text})$ extracts text from an image, $\textsf{TxtQA}_{\varphi}: \textsf{Text} \to (\texttt{answer}{:}\textsf{Text}, \texttt{Score}{:}\textsf{Float})$ answers a parameterized question $\varphi$ over text, and $\textsf{ImgCls}: \textsf{Image} \to (\texttt{label}{:}\textsf{Text})$ classifies an image.} \revision{Since tasks may be stochastic, $\llbracket t \rrbracket$ induces a distribution over output relations $\Delta(\mathsf{Rel}_{\mathbb{T}})$; we accordingly define plan equivalence \emph{in distribution} to state transformation correctness.}

A \emph{task-specific operator} correlates $t$ with each tuple of an input relation $r$ by binding $t$ to attributes $\bar{B}$, in the table-valued LATERAL/\textsf{Apply} (CROSS APPLY) style:
\[
\mathsf{Apply}_{t(\bar{B})}(r)\;=\;\biguplus_{x \in \mathrm{occ}(r)}\; \{x\}\times t(x[\bar{B}]).
\]
\revision{For example, $\mathsf{Apply}_{\textsf{OCR}(\texttt{img})}(r)$ extracts text from each \texttt{img} and appends it as a new attribute \texttt{txt}, yielding a relation with schema $(\texttt{pid}{:}\textsf{Int}, \texttt{img}{:}\textsf{Image}, \texttt{txt}{:}\textsf{Text})$.}
For readability, we depict this in pipeline form (e.g., $r \rightarrow \textsf{Op}$) as shorthand for $\mathsf{Apply}_{t(\bar{B})}(r)$ with the declared bindings. Semantics-dependent predicates are expressed by composing with \revision{relational} operators, e.g., $\sigma_{\psi}\!\left(\mathsf{Apply}_{t(\bar{B})}(r)\right)$, where $\psi$ is a well-typed predicate over $\mathbb{D}_{\mathrm{syn}}$ output attributes.

\para{Example (Decomposing an Image Claim Filter).}
Let $r(\texttt{pid}:\textsf{Int},\texttt{img}:\textsf{Image})$ be a relation of products.
Consider an intent $\mathcal{I}$: \emph{``Flag products whose package label contains compliance-sensitive claims (e.g., `FDA approved').''}
\revision{We realize this intent by composing $\textsf{OCR}$ and $\textsf{TxtQA}_{\varphi}$ introduced above, with $\varphi$ asking whether the extracted text contains a compliance claim.}
A \TxRA{} plan for the semantic filter is (omitting a final projection):
\begin{multline*}
\sigma_{\texttt{answer}=\text{``yes''}\wedge \texttt{Score}\ge \lambda}\!\Bigl(
\mathsf{Apply}_{\textsf{TxtQA}_{\varphi}(\texttt{txt})}\! \bigl(
\mathsf{Apply}_{\textsf{OCR}(\texttt{img})}(r)
\bigr)\Bigr).
\end{multline*}
Here, the intermediate OCR outputs become ordinary attributes that are passed to $\textsf{TxtQA}_{\varphi}$, while the final decision is expressed as a standard relational filter over $(\texttt{answer}, \texttt{Score})$.

\para{Conservativity.}
If we restrict $\mathbb{D}_{\mathrm{sem}}=\emptyset$ and $\Omega_{\mathrm{task}}=\emptyset$, then \TxRA{} reduces to classical relational algebra over $\mathbb{D}_{\mathrm{syn}}$.

\para{Algebraic Properties and Reorderability.}
We model tasks as side-effect-free, tuple-wise computations that append attributes via $\mathsf{Apply}$.
A sufficient condition to reorder two task-specific operators $Op_1$ and $Op_2$ is that neither consumes attributes produced by the other.
Let $\mathrm{Attr}(Op)$ be attributes generated by $Op$ and $\mathrm{Free}(Op)$ be required input attributes; then a safe commutativity condition is:
\[
\mathrm{Free}(Op_1) \cap \mathrm{Attr}(Op_2) = \emptyset \;\wedge\; \mathrm{Free}(Op_2) \cap \mathrm{Attr}(Op_1) = \emptyset .
\]
Similarly, a filter $\sigma_{\psi}$ can be pushed below $\mathsf{Apply}$ whenever $\psi$ does not reference attributes produced by the task. 

\revision{For example, $\mathsf{Apply}_{\textsf{OCR}(\texttt{img})}$ 
and $\mathsf{Apply}_{\textsf{TxtQA}_{\varphi}(\texttt{txt})}$ cannot be 
reordered since $\mathrm{Free}(\textsf{TxtQA}) \cap \mathrm{Attr}(\textsf{OCR}) 
= \{\texttt{txt}\}$, whereas an independent 
$\mathsf{Apply}_{\textsf{ImgCls}(\texttt{img})}$ shares no attribute 
dependency and swaps freely. Likewise, 
$\sigma_{\texttt{answer}{=}\text{``yes''}\wedge\texttt{Score}\ge\lambda}$ 
cannot be pushed below $\mathsf{Apply}_{\textsf{TxtQA}_{\varphi}}$ as it 
references its outputs, but a syntactic 
$\sigma_{\mathrm{filesize}(\texttt{img}) > X}$ pushes below all 
$\mathsf{Apply}$'s.}

\section{Logical Planner}\label{sec:logical_planner}

\begin{figure}[t]
\centering
\begin{tikzpicture}[
  font=\small\sffamily,
  actor/.style={draw, rectangle, rounded corners=1pt, minimum width=1.7cm, minimum height=0.45cm, fill=white, font=\bfseries\small\sffamily, inner sep=2pt},
  msg/.style={->, >=stealth, semithick},
  ret/.style={->, >=stealth, dashed, semithick},
  loopop/.style={->, >=stealth, semithick},
  exec/.style={draw, fill=white, line width=0.4pt, minimum width=0.18cm, inner sep=0pt},
  phaselbl/.style={font=\itshape\small\sffamily}
]
  \def\xUser{0}
  \def\xPlanner{3.4}
  \def\xLLM{6.8}
  \def\yBot{-3.60}
  % Phase background bands
  \fill[gray!12]  (-1.05,-0.50) rectangle (7.75,-2.20);
  \fill[green!12] (-1.05,-2.20) rectangle (7.75,-3.50);
  % Actor headers
  \node[actor] (user)    at (\xUser,0)    {Caller};
  \node[actor] (planner) at (\xPlanner,0) {Logical Planner};
  \node[actor] (llm)     at (\xLLM,0)     {LLM};
  % Lifelines
  \draw[dashed, gray!70] (\xUser,-0.3)    -- (\xUser,\yBot);
  \draw[dashed, gray!70] (\xPlanner,-0.3) -- (\xPlanner,\yBot);
  \draw[dashed, gray!70] (\xLLM,-0.3)     -- (\xLLM,\yBot);
  % Phase labels
  \node[phaselbl, rotate=90, align=center] at (-0.5,-1.35) {Seed Plan\\Synthesis};
  \node[phaselbl, rotate=90, align=center] at (-0.5,-2.85) {Plan\\Exploration};
  % Execution boxes
  \node[exec, minimum height=0.2cm] at (\xUser,-0.65) {};
  \node[exec, minimum height=0.2cm] at (\xUser,-3.30) {};
  \node[exec, minimum height=2.75cm, anchor=north] at (\xPlanner,-0.55) {};
  \node[exec, minimum height=1.20cm, anchor=north] at (\xLLM,-0.95) {};
  % Phase 1: Seed Plan Synthesis
  \draw[msg] (\xUser,-0.65)    -- node[above, midway, font=\footnotesize\sffamily] {intent + I/O schemas}      (\xPlanner,-0.65);
  \draw[msg] (\xPlanner,-1.05) -- node[above, midway, font=\footnotesize\sffamily, yshift=-2pt] {intent + operator catalog} (\xLLM,-1.05);
  \draw[ret] (\xLLM,-1.40)     -- node[above, midway, font=\footnotesize\sffamily, yshift=-2pt] {draft seed plan(s)}        (\xPlanner,-1.40);
  \draw[msg] (\xPlanner,-1.75) -- node[above, midway, font=\footnotesize\sffamily, yshift=-2pt] {critique \& refine prompt} (\xLLM,-1.75);
  \draw[ret] (\xLLM,-2.10)     -- node[above, midway, font=\footnotesize\sffamily, yshift=-2pt] {refined seed plan(s)}      (\xPlanner,-2.10);
  % Phase 2: Plan Exploration (self-loops)
  \draw[loopop]
    (\xPlanner+0.1,-2.40) -- ++(0.5,0) -- ++(0,-0.3) -- (\xPlanner+0.1,-2.70);
  \node[font=\footnotesize\sffamily, anchor=west] at (\xPlanner+0.6,-2.55) {structural rewrites};
  \draw[loopop]
    (\xPlanner+0.1,-2.85) -- ++(0.5,0) -- ++(0,-0.3) -- (\xPlanner+0.1,-3.15);
  \node[font=\footnotesize\sffamily, anchor=west] at (\xPlanner+0.6,-3.00) {semantic alternatives};
  % Output back to caller
  \draw[ret] (\xPlanner,-3.30) -- node[above, midway, font=\footnotesize\sffamily] {logical plans} (\xUser,-3.30);
\end{tikzpicture}
\vspace{-6mm}
\caption{\revision{The workflow of \OURSYSTEM{}'s logical planner: seed synthesis (gray) and plan exploration (green).}}
\vspace{-5mm}
\label{fig:logical_workflow}
\end{figure}
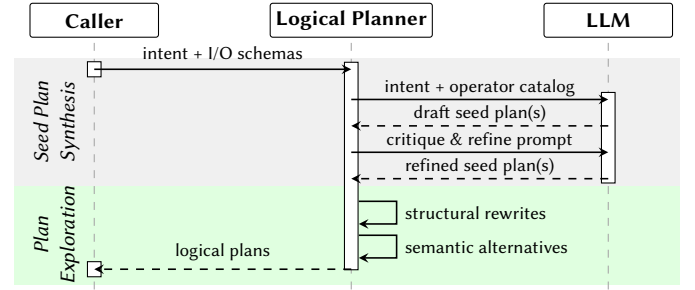

In this section, we present the logical planner. We first describe our task-specific logical operators (\S\ref{sec:logical_planner:ops}), then present seed plan synthesis (\S\ref{sec:seed_synthesis}), and finally describe plan exploration via \revision{structural rewrites and semantics-guided alternative generation} (\S\ref{sec:exploration}). \revision{Figure~\ref{fig:logical_workflow} summarizes the workflow.} We tested the logical planner on TPC-H queries specified as intents with I/O schemas, and found that it recovers the required relational structure---including multi-way joins---without introducing unnecessary operators.

\begin{table*}[t]
\centering
    \caption{Currently supported task-specific logical operators for text and image.} 
    \vspace{-3mm}
\label{tab:operator_ontology}
\small
\renewcommand{\arraystretch}{1.1}
\resizebox{\textwidth}{!}{%
\begin{tabular}{lllll}
\toprule
\textbf{Modality} & \textbf{Category} & \textbf{Operator} & \textbf{Output Schema} & \textbf{Description} \\ \midrule

% =========================
% TEXT MODALITY
% =========================
\multirow{17}{*}{\textbf{Text}} 
  & \multirow{3}{*}{\textbf{Structural}} 
  & \textbf{OpTxtSplit} 
  & $R_{Spl}(\text{SegIdx}, \text{Content})$ 
  & Segments text into ordered units  (e.g., sentences) for indexing. \\
  
  & & \textbf{OpTxtOutline} 
  & $R_{Out}(\text{SecIdx}, \text{ParentIdx}, \text{Level}, \text{Title}, \text{Content})$ 
  & Extracts hierarchical sections and their logical structure. \\
  
  & & \textbf{OpTxtTable} 
  & $R_{Tab}(\text{TabIdx}, \text{RowIdx}, \text{ColName}, \text{CellValue})$ 
  & Deconstructs embedded tables into relational rows and columns. \\
  \cmidrule(l){2-5}

  & \multirow{5}{*}{\textbf{Attributive}} 
  & \textbf{OpTxtNER} 
  & $R_{NER}(\text{Entity}, \text{Score}, \text{Word})$ 
  & Identifies and classifies entity spans. \\
  
  & & \textbf{OpTxtCls} 
  & $R_{Cls}(\text{Label}, \text{Score})$ 
  & Predicts labels for a text from a given label set with scores. \\
  
  & & \textbf{OpTxtSenti} 
  & $R_{Snt}(\text{Polarity}, \text{Score})$
  & Identifies sentiment polarity per aspect. \\
  
  & & \textbf{OpTxtQA} 
  & $R_{QA}(\text{Answer}, \text{Score})$ 
  & Extracts the answer value with its confidence score \\
  
  & & \textbf{OpTxtClaim} 
  & $R_{Clm}(\text{ClaimSpan}, \text{Verdict}, \text{EvdSpan}, \text{Score})$
  & Identifies claim spans and validates factuality with evidence. \\
  \cmidrule(l){2-5}

  & \multirow{5}{*}{\textbf{Associative}} 
  & \textbf{OpTxtRel} 
  & $R_{Rel}(\text{Subject}, \text{Predicate}, \text{Object}, \text{Score})$
  & Extracts semantic triples representing factual relationships. \\
  
  & & \textbf{OpTxtEvent} 
  & $R_{Evt}(\text{EvtIdx}, \text{Type}, \text{Role}, \text{Argument}, \text{Score})$
  & Extracts event frames as role--argument bindings. \\
  
  & & \textbf{OpTxtNLI}
  & $R_{NLI}(\text{Label}, \text{Score})$
  & Determines the logical inference relationship between two texts.  \\

  & & \textbf{OpTxtCoref} 
  & $R_{Ref}(\text{ClusterIdx}, \text{Span}, \text{Mention})$
  & Groups mentions referencing the same entity into clusters. \\

  & & \textbf{OpTxtPairScore} 
  & $R_{Sim}(\text{Score})$
  & Quantifies semantic similarity or relevance between a text pair. \\
\cmidrule(l){2-5}

  & \multirow{3}{*}{\textbf{Generative}} 
  & \textbf{OpTxtSum} 
  & $R_{Sum}(\text{Summary})$ 
  & Condenses the input text into a concise summary. \\
  
  & & \textbf{OpTxtTrans} 
  & $R_{Trn}(\text{Translation})$
  & Converts the input text into a specified target language. \\
  
  & & \textbf{OpTxtRewrite} 
  & $R_{Rew}(\text{Content})$ 
  & Paraphrases or modifies text style while preserving meaning. \\ \cmidrule(l){2-5}

  & \textbf{Latent} 
  & \textbf{OpTxtEmbed} 
  & $R_{Vec}(\text{Vector})$ 
  & Encodes text into a vector representation.
  \\ \midrule

% =========================
% IMAGE MODALITY
% =========================
\multirow{11}{*}{\textbf{Image}} 
  & \multirow{2}{*}{\textbf{Structural}} 
  & \textbf{OpImgRegion} 
  & $R_{Reg}(\text{RegIdx}, \text{BBox}, \text{Mask})$ 
  & Segments the image into indexed regions (boxes or masks). \\
  
  & & \textbf{OpImgOCR} 
  & $R_{OCR}(\text{Text}, \text{Score})$
  & Extracts text content with spatial coordinates. 
  \\ \cmidrule(l){2-5}

  & \multirow{4}{*}{\textbf{Attributive}} 
  & \textbf{OpImgCls} 
  & $R_{Cls}(\text{Label}, \text{Score})$ 
  & Classifies the entire image into categories. \\
  
  & & \textbf{OpImgObj} 
  & $R_{Obj}(\text{BBox}, \text{Label}, \text{Score})$ 
  & Detects objects with bounding boxes and class labels. \\
  
  & & \textbf{OpImgKeypt} 
  & $R_{Kpt}(\text{KptIdx}, \text{Point}, \text{Label})$ 
  & Identifies semantic keypoints (e.g., joints, landmarks). \\

  & & \textbf{OpImgVQA} 
  & $R_{VQA}(\text{Answer}, \text{Score})$ 
  & Answers natural language questions about the image. \\
  \cmidrule(l){2-5}

  & \multirow{2}{*}{\textbf{Associative}} 
  & \textbf{OpImgSceneRel} 
  & $R_{SGG}(\text{Subject}, \text{Predicate}, \text{Object}, \text{Score})$ 
  & Extracts visual relationships (scene graph triples). \\
  
  & & \textbf{OpImgPairScore} 
  & $R_{Sim}(\text{Score})$
  & Quantifies visual similarity between an image pair. \\
  \cmidrule(l){2-5}

  & \multirow{2}{*}{\textbf{Generative}} 
  & \textbf{OpImgCap} 
  & $R_{Cap}(\text{Caption})$ 
  & Generates a descriptive caption for the image. \\
  
  & & \textbf{OpImgEdit} 
  & $R_{Edit}(\text{Image})$ 
  & Generates a modified image based on constraints. \\
  \cmidrule(l){2-5}

  & \textbf{Latent} 
  & \textbf{OpImgEmbed} 
  & $R_{Vec}(\text{Vector})$ 
  & Encodes the image into a dense vector representation.
    \\
\bottomrule
\end{tabular}%
}
\vspace{-4mm}
\end{table*}

\subsection{Task-Specific Logical Operators} \label{sec:logical_planner:ops}

Task-specific operators map \revision{unstructured} inputs to structured relations that can be processed by \revision{relational} operators. We address two design questions: (i) the granularity of task-specific operators, and (ii) how to organize the operator set in a database-centric way.

\para{Granularity.} We define each task-specific logical operator at the level of a common \emph{task formulation} in standard ML model ecosystems (e.g., Hugging Face): a reusable computation with a typed signature and a stable output schema independent of any backend. \revision{The catalog is task-family based rather than workload-specific. Each operator corresponds to a common ML task formulation in standard model ecosystems such as Hugging Face Tasks. For example, \textsf{OpTxtSenti} specializes text classification with a sentiment label space, rather than encoding a Movie scenario--specific rule.}

\para{Task Categories.}
We group task-specific operators by the \emph{database role} of their outputs into five categories:
\emph{Structural} (indexes substructures such as segments/regions),
\emph{Attributive} (unary descriptors such as labels/answers),
\emph{Associative} (relations over elements or pairs, e.g., triples/NLI),
\emph{Generative} (derived artifacts as new semantic objects), and
\emph{Latent} (vectors for similarity-based retrieval).

\para{Current Operator Set and Extensibility.}
Table~\ref{tab:operator_ontology} lists the supported text and image operators. New operators are added by registering a task signature and assigning a category.

\subsection{Seed Plan Synthesis} \label{sec:seed_synthesis}

Given a semantic operator instance, the planner performs up to $B_{log}$ synthesis attempts; each attempt runs \emph{draft plan synthesis} followed by \emph{plan refinement} and yields at most one seed plan. To keep LLM usage budgeted, refinement is capped by $b_{\mathrm{ref}}$ LLM-invoking verification/repair rounds, while deterministic checks do not count.

% \begin{figure}[t]
% \centering
% \includegraphics[width=\linewidth]{figures/logical.pdf}
% \vspace{-6mm}
% \caption{Workflow of the seed plan synthesis. Drafts may include non-catalog placeholders, which refinement grounds to equivalent catalog subgraphs when possible.}
% \vspace{-5mm}
% \label{fig:logical_workflow}
% \end{figure}

\subsubsection{Draft Plan Synthesis}
Each attempt begins by generating a \emph{draft} seed plan that outlines a plausible sequence of reasoning steps for the instance.
The planner prompts an LLM with the intent, input/output schemas, the operator catalog, and a short history of previous plans to discourage repetition and encourage structural diversity (e.g., after a VQA-centric draft, a later attempt may explore an embedding-based alternative). \jhha{Here, extensibility is a core design principle of \OURSYSTEM{}; since the planner uses in-context learning over the prompt-provided operator catalog, adding a new modality (e.g., video) or a domain-specific operator typically only requires registering its signature and description in the catalog, without fine-tuning or retraining.}
While the catalog provides the primary vocabulary, drafts may include non-catalog placeholders when needed; each placeholder must declare a typed signature and explicit output schema to remain type-checkable. 

\subsubsection{Plan Refinement}
Drafts may be ill-formed (e.g., unsupported operators, invalid wiring, or schema mismatches). The planner refines each draft via a compile-time fix loop: it runs a deterministic verifier to collect all errors and return a concrete report, prompts an LLM to revise the draft to satisfy the report, and repeats until the verifier passes or the refinement budget is exhausted. Once verifier-clean, the planner performs \emph{catalog grounding}, rewriting non-catalog placeholders or composite steps into equivalent subgraphs of catalog primitives when possible. For example, if a filter predicate references a non-existent attribute (\texttt{Ans}), the verifier reports the missing field and the revision updates the predicate to use the correct upstream output attribute. Similarly, the placeholder \texttt{OpTxtImgSim} is grounded into \texttt{OpImgEmbed}$\rightarrow$\texttt{OpTxtEmbed}.

\subsection{Plan Exploration} \label{sec:exploration}

\revision{Plan exploration expands each seed plan via two mechanisms. Structural rewrites follow the classical optimizer model: correctness-preserving transformations whose preconditions are checked from plan wiring, I/O bindings, and operator contracts. Semantic alternatives exploit task-level knowledge (e.g., partitionability, modality conversion) to propose additional plans without assuming equivalence, and are retained only if their compiled physical 
plans improve the quality--latency--cost trade-off on the validation/teacher signal.}

\subsubsection{\revision{Structural Rewrites}}
\revision{Structural rewrites are the only plan-exploration steps that we treat as correctness-preserving algebraic rewrites. They} require only (i) data dependencies (I/O bindings) and (ii) an operator contract over schema and cardinality.
In \TxRA{}, each task-specific operator is a per-tuple \textsf{Apply} that appends a known output schema, thus we apply classical transformations whenever the dependency conditions from \S\ref{sec:algebra} hold.
Concretely, two operators commute (and can be reordered) whenever
$\mathrm{Free}(Op_1)\cap \mathrm{Attr}(Op_2)=\emptyset$ and
$\mathrm{Free}(Op_2)\cap \mathrm{Attr}(Op_1)=\emptyset$. A filter $\sigma_{\psi}$ can be pushed below an \textsf{Apply} when $\psi$ does not reference task-produced attributes.
Our prototype implements two rules: \textsc{FilterPushdown} and \textsc{CommutativeReordering}.

\subsubsection{\revision{Semantics-Guided Alternative Generation}}
\revision{Beyond structural rewrites, useful alternatives often require task-level properties that are not implied by schema or cardinality contracts. We therefore do not present semantic alternatives as equivalence rules. Instead, each semantic alternative is generated by an alternative-generation template: it proposes a new typed DAG that may be cheaper or more accurate for a particular intent/data distribution, and the optimizer decides whether to keep it through validation/teacher-based plan ranking. Some templates can be equivalence-preserving under additional task-specific assumptions, such as lossless partitioning, but \OURSYSTEM{} does not rely on that assumption for correctness; it treats the generated plan as an alternative. Our prototype uses two high-impact templates for unstructured workloads.}

\para{\revision{Template 1: Partitioned Extraction.}}
\revision{Some task-specific operators admit useful partitioned alternatives: the input is split into subrelations by a partitioner $P$, the task is applied to each partition, and the outputs are normalized and unioned. This template is equivalence-preserving only when the task is partition-local and the normalizer preserves all cross-partition identifiers needed by downstream operators. Otherwise, it is an approximate alternative that trades possible context loss for lower latency/cost:}
\begin{equation}
  \mathsf{Apply}_{t}(r)\;\leadsto\;
  \mathsf{Norm}\!\left(\biguplus_{s \in P(r)} \mathsf{Apply}_{t}(s)\right).
\end{equation}
\emph{Example.} For entity extraction over a long PDF, the planner splits into pages, runs extraction per page in parallel, and normalizes span offsets before unioning results.
\emph{Benefit.} Enables intra-operator parallelism and mitigates context-window limits.
\revision{\emph{Validation role.} Because cross-partition context may matter, the resulting plan competes with non-partitioned plans and is selected only when it improves the measured quality--latency--cost trade-off.}

\para{\revision{Template 2: Cross-Modal Proxying.}}
\revision{Some tasks admit cheaper alternatives by converting modalities through a conversion operator $c$, and then applying the task in the converted modality:}
\begin{equation}
  \mathsf{Apply}_{t_{\mathrm{img}}}(r)\;\leadsto\;
  \mathsf{Apply}_{t_{\mathrm{txt}}}\!\left(\mathsf{Apply}_{c}(r)\right).
\end{equation}
\revision{This alternative is not assumed to be semantically equivalent to the original image task; it is a proxy whose usefulness depends on the intent and data distribution.}
\emph{Example.} To filter images containing a specific phrase, the planner applies OCR and then runs text filtering/classification over the extracted strings, instead of invoking a heavy vision model.
\emph{Benefit.} Enables modality switching and cheaper proxy execution when high-fidelity models are costly or unavailable.
\revision{\emph{Safeguard.} Because conversion can fail, \OURSYSTEM{} can compile this template with a physical fallback, e.g., routing empty or low-confidence OCR outputs to a stronger vision backend. The final decision is made by the same validation/teacher-based plan ranking used for other alternatives.}
\section{Physical Planner} \label{sec:physical_planner}

In this section, we present the physical planner. We first describe the physical plan construction (\S\ref{sec:physical_planner:con}), focusing on task-specific operators that compose family-specific implementations with a data-aware router. We then describe multi-objective tuning (\S\ref{sec:tuning}), where Bayesian optimization tunes parameters on a validation set.

\subsection{Physical Plan Construction} \label{sec:physical_planner:con}

Algorithm~\ref{alg:parametric_compilation} constructs a physical plan $\mathcal{P}$ and its tunable parameter space $\Theta$ from a logical plan $\mathcal{L}$.
It initializes an empty plan and parameter space (Line~\ref{ln:pp:init}) and traverses operators in topological order (Line~\ref{ln:pp:toposort}).
For each task-specific operator (Line~\ref{ln:pp:taskspec}), the compiler considers each synthesis directive $d\in\mathcal{D}$ (Line~\ref{ln:pp:loopS}). The implementation catalog indexes entries by both the operator $l$ and the directive $d$ (i.e., backend family/style), so lookups are keyed by $(d,l)$ (Line~\ref{ln:pp:lookup}); otherwise, it synthesizes the missing implementation using the directive and the backend registry, which records available model-serving resources for task-specific inference and their instantiation interfaces (Line~\ref{ln:pp:synth}).
It collects candidate implementations and their exposed parameters (Lines~\ref{ln:pp:addcand}--\ref{ln:pp:addparams}), then synthesizes a data-aware router (Line~\ref{ln:pp:router}) and forms a composite operator (Line~\ref{ln:pp:spcf}).
For each \revision{relational} operator (Line~\ref{ln:pp:agnostic}), it binds a concrete implementation from the catalog and exposes its tunable parameters (Line~\ref{ln:pp:agn}).
Each bound operator is added to $\mathcal{P}$ (Line~\ref{ln:pp:addnode}) and its parameters are aggregated into $\Theta$ (Line~\ref{ln:pp:aggOmega}).
Finally, the compiler copies the dataflow edges from $\mathcal{L}$ to $\mathcal{P}$ (Line~\ref{ln:pp:link}) and returns $\mathcal{P}$ and $\Theta$ (Line~\ref{ln:pp:return}).

\begin{algorithm}
\caption{Physical Plan Construction}
\label{alg:parametric_compilation}
\begin{algorithmic}[1]
\Statex \textbf{Input:} $\mathcal{L}$ (logical plan DAG), $\mathcal{C}$ (implementation catalog), $\mathcal{D}$ (synthesis directives), $\mathcal{R}$ (backend registry)
\Statex \textbf{Output:} $\mathcal{P}$ (physical plan), $\Theta$ (tunable parameter space)
\Statex \hrulefill

\State $\mathcal{P} \leftarrow \textsc{EmptyDAG}(); \ \Theta \leftarrow \emptyset$ \label{ln:pp:init}

\For{\textbf{each} $l \in \textsc{TopologicalSort}(\mathcal{L})$} \label{ln:pp:toposort}
  \State $\Theta_l \leftarrow \emptyset$ \label{ln:pp:initOmegal}

  \If{$\textsc{IsTaskSpecific}(l)$} \label{ln:pp:taskspec}
    \State $\mathcal{K} \leftarrow \emptyset$ \label{ln:pp:initK}
    \For{\textbf{each} $d \in \mathcal{D}$} \label{ln:pp:loopS}
      \State $I \leftarrow \textsc{LookupCatalog}(\mathcal{C}, d, l)$ \label{ln:pp:lookup}
      \If{$I = \varnothing$} \label{ln:pp:miss}
        \State $I \leftarrow \textsc{SynthesizeImpl}(d, l, \mathcal{R})$ \label{ln:pp:synth}
      \EndIf
      \If{$I \neq \varnothing$} \label{ln:pp:hit}
        \State $\mathcal{K} \leftarrow \mathcal{K} \cup \{I\}$ \label{ln:pp:addcand}
        \State $\Theta_l \leftarrow \Theta_l \cup \textsc{Params}(I)$ \label{ln:pp:addparams}
      \EndIf
    \EndFor
    \State $r \leftarrow \textsc{SynthesizeRouter}(\mathcal{K})$ \label{ln:pp:router}
    \State $\Theta_l \leftarrow \Theta_l \cup \textsc{Params}(r)$ \label{ln:pp:routerparams}
    \State $p_{\mathrm{spcf}} \leftarrow (r,\mathcal{K}); \ \ p \leftarrow p_{\mathrm{spcf}}$ \label{ln:pp:spcf}

  \Else \label{ln:pp:agnostic}
    \State $(p_{\mathrm{agn}},\Theta_l) \leftarrow \textsc{\revision{BindRelationalOp}}(\mathcal{C}, l); \ \ p \leftarrow p_{\mathrm{agn}}$ \label{ln:pp:agn}
  \EndIf

  \State $\mathcal{P}.\textsc{AddNode}(p)$ \label{ln:pp:addnode}
  \State $\Theta \leftarrow \Theta \cup \Theta_l$ \label{ln:pp:aggOmega}
\EndFor

\State $\mathcal{P}.\textsc{LinkEdgesFrom}(\mathcal{L})$ \label{ln:pp:link}
\State \Return $\mathcal{P}, \Theta$ \label{ln:pp:return}
\end{algorithmic}
\end{algorithm}

% \begin{figure}[ht]
% \vspace{-2mm}
% \centering
% \includegraphics[width=\linewidth]{figures/physical.pdf}
% \vspace{-5mm}
% \caption{Workflow of the task-specific operator compilation.}
% \vspace{-5mm}
% \label{fig:physical_workflow}
% \end{figure}

\para{\textsc{SynthesizeImpl} (Line~\ref{ln:pp:synth}).}
It materializes a concrete physical implementation for a task-specific logical operator $l$.
It takes a synthesis directive $d\in\mathcal{D}$---a prompt template that targets a backend family (symbolic, specialized, general-purpose, or composite) and encodes the intended implementation style (e.g., ``use only foundation models'' or ``compose a multi-step pipeline''). Concretely, the planner instantiates this template with $(d, l, \mathcal{R})$—the directive $d$, the operator specification of $l$ (typed signature, input bindings, and output schema), and the backend registry $\mathcal{R}$, including only entries applicable to $d$ and $l$ to keep prompts compact. The LLM returns a callable implementation that satisfies the operator contract and declares its tunable parameters (e.g., confidence thresholds, top-$k$), which are added to the plan's parameter space for tuning. For example, consider \textsf{OpTxtQA}. The planner synthesizes one candidate per backend family---a symbolic regex filter (no parameters), a specialized \textsc{DistillBERT\_QA} variant with threshold $\theta_{\mathrm{qa}}$, a \textsc{GPT4}-based variant (no parameters), and a composite that prefilters with \textsc{CLIPSim} and escalates to \textsc{GPT4} with thresholds $\theta_{\mathrm{sim}}$ and $\theta_{\mathrm{gpt}}$. 
Since synthesis may fail, \textsc{SynthesizeImpl} is paired with \S\ref{sec:impl:robust}.

\para{\textsc{SynthesizeRouter} (Line~\ref{ln:pp:router}).}
Given candidate implementations $\mathcal{K}=\{I_1,\ldots,I_N\}$, \textsc{SynthesizeRouter} constructs a data-aware router $r$ that dispatches each input to an implementation based on its estimated difficulty. The router consists of: (i) \emph{lightweight input-feature extraction} code, (ii) \emph{scoring} code that maps these features to a scalar difficulty score $s\in[0,1]$ (higher means ``needs more semantic capability''), and (iii) \emph{partitioning/dispatching} code. (i) and (ii) are synthesized by prompting an LLM with the operator specification, the ordered implementation set $\mathcal{K}$, and a router generation prompt. (iii) is then generated by instantiating a fixed bucketing-and-dispatch template which uses the scorer and $N\!-\!1$ tunable cutpoints to assign each row to one of $N$ buckets and invoke the corresponding $I_i$. The features are cheap, non-inferential signals (e.g., text length/format; image resolution/aspect ratio, and simple pixel statistics). \textsc{SynthesizeRouter} is also paired with \S\ref{sec:impl:robust}.

\para{Percentile Bucketing with Softmax-Parameterized Masses.}
In practice, we bucket by \emph{percentiles} rather than raw score thresholds, making routing depend on the relative ordering rather than absolute score values, which makes it less sensitive to score-distribution shifts. Concretely, we convert scores to rank-based percentiles $u\in(0,1]$ and assign buckets using $N\!-\!1$ increasing percentile cutpoints $\tau$ via $k=\mathrm{digitize}(u;\tau)$.
To avoid ordered constraints on $\tau$, we parameterize bucket masses with logits $\alpha\in\mathbb{R}^{N}$, define $\pi=\mathrm{softmax}(\alpha)\in\Delta^{N-1}$, and set cutpoints by the CDF $\tau_k=\sum_{i=1}^{k}\pi_i$ for $k=1,\ldots,N-1$.

\para{Precompilation.} For the operators in Table~\ref{tab:operator_ontology}, we precompile physical implementations offline \revision{as a caching mechanism for efficiency}, except symbolic ones that require query-specific knowledge and are instantiated on demand. \revision{If an implementation is missing or the backend set changes, \OURSYSTEM{} synthesizes it at runtime.}

\subsection{Multi-Objective Tuning} \label{sec:tuning}

Given a compiled physical plan $\mathcal{P}$ with tunable parameters $\Theta$ and a validation set $\mathcal{E}$, the physical planner tunes $\mathcal{P}$ under a fixed evaluation budget $B_{\text{tune}}$, defined in terms of optimization trials (i.e., maximum plan executions) using Bayesian optimization. For $b=1,\ldots,B_{\text{tune}}$, BO proposes $\theta_b\in\Theta$; we execute $\mathcal{P}(\theta_b)$ on $\mathcal{E}$ and measure quality, latency, and cost. This tunes both router parameters and per-implementation parameters. As defined in \S\ref{sec:overview:moop}, we select the final plan by maximizing $U(\mathcal{P})$ with normalized latency and cost. We use log normalization since latency and cost span orders of magnitude: it compresses extremes and reduces sensitivity to outliers, budget caps, and cap choices.

\para{Tuning Practicalities.}
Handling high-dimensional mixed discrete--continuous search spaces is orthogonal to our contributions~\cite{turbo, SAASBO}. For efficiency, we cap each plan at 15 continuous parameters. To reduce sensitivity to initial random trials, we warm-start with a few hand-designed configurations spanning distinct backend regimes.
\section{Implementation} \label{sec:implementation}

In this section, we describe \OURSYSTEM{}’s implementation and deployment details. \OURSYSTEM{} is implemented in Python (\texttt{$\sim$20K} LoC) and currently supports text, image, and audio modalities.

\subsection{System Environment \& Stack} \label{sec:impl:stack}

\para{Core.}
\OURSYSTEM{} is implemented in Python (3.10+) and uses PyTorch (2.6+).
For BO, we use BoTorch via Optuna's BoTorch integration.

\para{Integration.}
\revision{\OURSYSTEM{} is a standalone optimizer/executor with its own CLI: it can run through its own CLI without depending on any host SQPE. To interoperate with existing SQPEs, the host SQPE implements a small adapter: (i) a \emph{dispatch hook} forwards the host's semantic operator calls to \OURSYSTEM{}, (ii) \emph{I/O adapters} exchange relations as \texttt{pandas.DataFrame}, and (iii) an \emph{LM adapter} routes \OURSYSTEM{}'s LM calls through the host's LM client; \OURSYSTEM{} itself remains unchanged. Our experiments use a LOTUS integration; the same pattern applies to other SQPEs.}

\para{Data Model.}
Operator inputs/outputs are passed through the adapter as Pandas DataFrames (Arrow-compatible when applicable),
enabling vectorized execution and efficient batch inference (e.g., batching LLM or local model calls) rather than per-tuple invocation.

\subsection{Optimization Overhead Management} \label{sec:impl:overhead}

\para{Plan Caching.}
To amortize optimization cost, \OURSYSTEM{} caches optimized physical plans keyed by an \emph{intent signature} that captures the intent-defining configuration.
On a cache hit, \OURSYSTEM{} reuses the cached physical plan without re-optimizing.
\jhha{To handle data distribution drift, which can degrade synthesized symbolic rules and routers, we use two strategies. First, percentile bucketing relies on relative ranks and is thus less sensitive to minor score shifts. Second, cached plans are considered as valid only for a specific data profile. In production, we recommend periodic re-optimization to regenerate symbolic rules and re-tune parameters on fresh samples.}

\para{Adaptive Bypass.}
For small inputs where optimization overhead can dominate, \OURSYSTEM{} bypasses optimization and runs a default LLM-only execution.
We use a simple cardinality heuristic $N \le \tau$.
\jhha{We set $\tau{=}300$ from profiling the crossover between LLM-only and \OURSYSTEM{} execution; near this scale, only \revision{16/55} queries favored LLM-only, so bypassing is a conservative guardrail and $\tau$ is configurable.}

\subsection{Robustness} \label{sec:impl:robust}

\para{Code Safety.}
LLM-synthesized code includes a small test wrapper (expected signature and required columns) and is smoke-tested on mock inputs. It then runs in an isolated runtime with static checks and timeouts to prevent unsafe operations and bound resource usage. If synthesized code fails the smoke test, \OURSYSTEM{} re-synthesizes it under a bounded timeout/attempt budget until it passes.

\para{Error Recovery.} \jhha{At runtime, \OURSYSTEM{} handles transient failures with bounded retries and safe-mode fallback. If a synthesized code raises an exception or returns a confidence score below a safety margin (when available), we fall back to a default LLM to ensure completion even in the presence of synthesized code fragility.}
\section{Experiments} \label{sec:experiments}

 We seek to answer the following research questions about \OURSYSTEM{}:

\begin{list}{}{%
    \setlength{\leftmargin}{2.5em}
    \setlength{\labelsep}{0.5em}
}
    \item[\textbf{RQ1.}] Does \OURSYSTEM{} achieve better end-to-end quality/latency/cost trade-offs than prior SQPE optimizers? (\S\ref{sec:endtoendperf})

    \item[\textbf{RQ2.}] How sensitive is \OURSYSTEM{} to its configurations? (\S\ref{sec:sensitivity})

    \item[\textbf{RQ3.}] Which techniques of \OURSYSTEM{} account for its performance gains, and by how much? (\S\ref{sec:ablation})

    \item[\textbf{RQ4.}] How does \OURSYSTEM{}'s performance scale with data size? (\S\ref{sec:scalability})

    \item[\revision{\textbf{RQ5.}}] \revision{How important is intent-specific synthesis? (\S\ref{sec:intent_specific_synthesis})}

    \item[\revision{\textbf{RQ6.}}] \revision{Does \OURSYSTEM{} outperform baselines under the same utility function across varied weights? (\S\ref{sec:appx:utility-sweep})}

    \item[\textbf{RQ7.}] Can \OURSYSTEM{} approximate the Pareto frontier? (\S\ref{sec:pareto})

\end{list}

\begin{table*}[ht]
\centering
\caption{Overview of datasets used in each scenario.}
\vspace{-3mm}
\label{tab:benchmark_overview}
\begin{tabular}{lclrl}
\toprule
\textbf{Scenario} & \textbf{SF} & \textbf{Table} & \textbf{\#Rows} & \revision{\textbf{Unstructured Columns Used}} \\
\midrule
Movie       & 20000  & Reviews                    & 20000 & ReviewText {\textbf{TEXT}} \\
\midrule
Wildlife    & 4000 & \begin{tabular}[t]{@{}l@{}}ImageData\\\revision{AudioData}\end{tabular}      & \begin{tabular}[t]{@{}r@{}}4000\\\revision{650}\end{tabular}      & \begin{tabular}[t]{@{}l@{}}Image {\textbf{IMAGE}}, City {\textbf{TEXT}}, StationID {\textbf{TEXT}}\\\revision{Audio {\textbf{AUDIO}}, City {\textbf{TEXT}}, StationID {\textbf{TEXT}}}\end{tabular} \\
\midrule
\revision{MMQA}        & \revision{300}    & \revision{\begin{tabular}[t]{@{}l@{}}LizzyCaplanText\\BenPiazzaText\\TampaInternationalAirport\\Images\end{tabular}} & \revision{\begin{tabular}[t]{@{}r@{}}300\\300\\300\\300\end{tabular}} & \revision{\begin{tabular}[t]{@{}l@{}}Title {\textbf{TEXT}}, Text {\textbf{TEXT}}\\Text {\textbf{TEXT}}\\Airlines {\textbf{TEXT}}, Destinations {\textbf{TEXT}}\\Image {\textbf{IMAGE}}\end{tabular}} \\
\midrule
\revision{Cars}        & \revision{100000} & \revision{\begin{tabular}[t]{@{}l@{}}CarComplaint\\CarImage\\CarAudio\end{tabular}} & \revision{\begin{tabular}[t]{@{}r@{}}99929\\19156\\871\end{tabular}} & \revision{\begin{tabular}[t]{@{}l@{}}Summary {\textbf{TEXT}}\\Image {\textbf{IMAGE}}\\Audio {\textbf{AUDIO}}\end{tabular}} \\
\midrule
E-Commerce  & 5000   & StyleDetails               & 5000  & \begin{tabular}[t]{@{}l@{}}ProductDisplayName {\textbf{TEXT}}, ProductDescriptors {\textbf{TEXT}}, Images {\textbf{IMAGE}}\end{tabular} \\
\bottomrule
\end{tabular}
\vspace{-2mm}
\end{table*}

\begin{table}[ht]
\caption{\revision{End-to-end average improvement of \OURSYSTEM{} over 
each baseline. $\Delta$ for Q is the arithmetic mean of $Q_{\OURSYSTEM{}} - Q_{\text{baseline}}$; $\times$ for L/C is the 
geometric mean of $\text{baseline}/\OURSYSTEM{}$. Queries unsupported or failed by a baseline are excluded; per-scenario raw values in 
Tables~\ref{tab:appx:movie}--\ref{tab:appx:ecomm}.}}
\label{tab:semantic_op_benchmark_summary}
\centering
\setlength{\tabcolsep}{6pt}
\renewcommand{\arraystretch}{1.2}
\vspace{-3mm}
\begin{tabular}{lccc}
\toprule
\textbf{System} & $\boldsymbol{\Delta}$ \textbf{Q} & $\boldsymbol{\times}$ \textbf{L} & $\boldsymbol{\times}$ \textbf{C} \\
\midrule
LOTUS              & +0.10 & 3.80$\times$  & 4.88$\times$  \\
Palimpzest         & +0.11 & 4.42$\times$  & 21.83$\times$ \\
ThalamusDB         & +0.28 & 4.33$\times$  & 1.51$\times$  \\
DocETL             & +0.08 & 13.19$\times$ & 8.62$\times$  \\
Extended AFlow     & +0.26 & 1.40$\times$  & 1.96$\times$  \\
Extended DyFlow    & +0.21 & 1.07$\times$  & 1.40$\times$  \\
\bottomrule
\end{tabular}
\end{table}

\begin{table}[t]
\centering
\small
\setlength{\tabcolsep}{6pt}
\renewcommand{\arraystretch}{1.12}
\caption{Fraction of total latency/cost spent in optimization (summing over optimization-enabled queries), and an amortized variant that reuses cached plans per scenario.}
\vspace{-3mm}
\label{tab:opt_overhead}
\begin{tabular}{lcccc}
\toprule
\textbf{Scenario} & \textbf{Overh.$L$\%} & \textbf{Amort.$L$\%} & \textbf{Overh.$C$\%} & \textbf{Amort.$C$\%} \\
\midrule
Movie    & 23.7 & 9.4 & 17.0 & 6.4 \\
Wildlife & 23.8 & 9.4 & 19.8 & 7.6 \\
\revision{MMQA}     & \revision{20.1} & \revision{7.7} & \revision{15.2} & \revision{5.6} \\
\revision{Cars}     & \revision{17.2} & \revision{5.9} & \revision{15.3} & \revision{4.5} \\
E-Comm   & 16.1 & 6.0 & 15.8 & 5.9 \\
\bottomrule
\end{tabular}
\vspace{-4mm}
\end{table}

\subsection{Experimental Setup}

\subsubsection{Benchmark} \label{sec:benchmarks}

We evaluate on SemBench~\cite{sembench}, a benchmark for SQPEs that features diverse semantic operators over \revision{text, image, and audio} data.\footnote{\revision{The Medical scenario was removed from SemBench due to licensing issues~\cite{sembench-medical-issue}.}} We enforce a uniform 24-hour per-query timeout. Table~\ref{tab:benchmark_overview} summarizes dataset properties.

\para{Movie.} This scenario evaluates sentiment-aware analytics over movie text reviews. It contains 10 semantic queries (Q1--Q4: filters; Q5--Q7: joins; Q8: classification; Q9--Q10: rankings).

\para{Wildlife.} \revision{This scenario studies species presence and co-occurrence in camera-trap images and animal sound audio captures from Kaggle datasets, with species labels available as ground truth. It contains 10 semantic queries: Q1/Q3/Q7/Q10 are image filters over species (with downstream aggregations), Q2/Q4/Q8 are audio filters, and Q5/Q6/Q9 are multi-modal queries combining image and audio.}

\para{\revision{MMQA.}} \revision{This scenario evaluates question answering over heterogeneous tables, text, and images. It contains 11 semantic queries: Q1/Q3a/Q3f/Q4/Q5 and Q6a--Q6c are text filters/maps, Q7 is a text$\to$image join, and Q2a--Q2b are image joins with an extraction.}

\para{\revision{Cars.}} \revision{This scenario combines vehicle metadata, NHTSA complaints, vehicle damage images, and engine-diagnostic audio. It contains 10 semantic queries: Q1/Q4/Q10 are text filters/maps, Q3/Q7/Q8 are image filters, Q2 is an audio filter, and Q5/Q6/Q9 are multi-modal queries combining text/image/audio.}

\para{E-Commerce.} This scenario uses an online fashion retail dataset with structured product attributes, text descriptions, and images. It contains \revision{14} semantic queries: Q1--Q2 are filters, Q3--Q4 maps, Q5--Q6 classifications, Q7--Q9 joins\revision{, Q10--Q11 are multi-stage 3-/4-way joins, Q12 a structured extraction, Q13 a cross-modal filter--join, and Q14 a filter with top-$k$ ranking}.

\subsubsection{Measure} \label{sec:measure}

We measure efficiency by latency (seconds) and monetary cost (USD), and measure quality in $[0,1]$ against ground truth: relative error (aggregation), F1 (retrieval), Spearman’s $\rho$ (ranking), and ARI (grouping/classification). We repeat each experiment three times and report geometric means. For brevity, we denote \textbf{Q} as quality ($\uparrow$), \textbf{L} as latency ($\downarrow$), and \textbf{C} as monetary cost ($\downarrow$).

\subsubsection{Environment}
All experiments ran on a machine with 2$\times$ RTX 3090 (24\,GB each), a 64-core Intel Xeon CPU, and 512\,GB RAM. {We accessed LLMs via Azure AI Foundry (no local GPUs) and report cost as Azure API charges plus estimated GPU cost for local models from runtime, using \$0.71/hour (NCasT4\_v3) as a proxy. We used Global Standard, disabled content filters, and set temperature to 0.

\subsubsection{\OURSYSTEM{} Settings} 

We use \GPTFOUR{} for planning (parallelism 20); for fair comparison, adaptive bypass uses \GPTFOURMINI{} (\revision{for audio inputs, \GPTAUDIOMINI{}}) and disable plan caching.
Unless otherwise noted, we use a 50-sample validation set, $B_{log}=3$, 20 BO trials, and set $w_{\text{q}}{=}1.0$, $w_{\text{l}}{=}0.4$, $w_{\text{c}}{=}0.3$.

\subsubsection{Competitors} \label{sec:competitors}

We compare against SQPEs with semantic operator optimization: Palimpzest~\cite{palimpzest} (with Abacus~\cite{abacus}), LOTUS~\cite{LOTUS}, ThalamusDB~\cite{thalamusdb}, and DocETL~\cite{DOCETL}. We also extend and evaluate AFlow~\cite{aflow} \revision{and DyFlow~\cite{dyflow}}. For all applicable systems, we set parallelism to 20. Following \cite{sembench}, we exclude AOP~\cite{AOP}, CAESURA~\cite{caesura}, and Unify~\cite{unify}, which require natural-language inputs. We use a controlled setting where all systems share the same pool of foundation models and, where supported, specialized backends. \revision{For audio inputs, every baseline uses \GPTAUDIOMINI{}.}

\para{LOTUS.} LOTUS~\cite{LOTUS} optimizes semantic operators by approximating an oracle LLM-based algorithm using proxy models. We use LOTUS v1.1.4 and set the oracle LLM to \GPTFOURMINI{} (\GPTFOUR{} was too slow for some queries). For semantic filters, we use \GPTFOURNANO{} as the proxy; for semantic joins, we use \texttt{e5-base-v2} (text) and \texttt{clip-ViT-B-32} (image) embeddings as proxies following~\cite{sembench}. \revision{For fair comparison on top-$k$ queries, we extend LOTUS with a Volcano-style pull executor that streams rows through semantic operators and cancels in-flight LLM calls once $k$ outputs are produced.}

\para{Palimpzest.} Palimpzest~\cite{palimpzest} integrates the Abacus, a cost-based optimizer~\cite{abacus}, which compares physical operators under a chosen objective and constraints with an LLM judge. We use Palimpzest v1.1.1 with Abacus enabled and set \texttt{MaxQualityAtFixedCost} with a \$10 constraint.\footnote{Abacus supports multiple objectives, including \texttt{MaxQuality}, \texttt{MinCost}, \texttt{MinLatency}, \texttt{MaxQualityAtFixedCost}, \texttt{MaxQualityAtFixedTime}, \texttt{MinCostAtFixedQuality}, and \texttt{MinTimeAtFixedQuality}. Without an explicit budget constraint, \texttt{MaxQuality} can select prohibitively expensive plans (e.g., $>$\$1000), which exceed our budget.} We provide \GPTFOURNANO{}, \GPTFOURMINI{}, \GPTFOUR{}, and \TEXTEMBEDDINGSMALL{} as candidates, with \GPTFOUR{} judge.

\para{ThalamusDB.}
ThalamusDB~\cite{thalamusdb} employs approximate query processing (AQP) with user-defined termination conditions (e.g., error bounds and token budgets), trading quality for efficiency via early stopping.
Because our evaluation targets near-exact outputs, enabling AQP termination can stop execution before meeting the required quality.
We therefore run ThalamusDB by setting the target error to zero and removing the token budget cap.
We use ThalamusDB v0.1.15 and fix the LLM to \GPTFOURMINI{}.

\para{DocETL.} DocETL~\cite{DOCETL} rewrites a  semantic operator pipeline with LLMs to improve quality only. We include DocETL v0.2.6 as a quality-oriented baseline, using \GPTFOURMINI{} as the default model and \TEXTEMBEDDINGSMALL{} for embeddings, and run via its Python API with optimization enabled. Since DocETL currently supports only text inputs, we evaluate it only on text-only queries and disable result caching for fairness. DocETL frequently fails at runtime \cite{LOTUS}; we report the average over three successful runs.

\para{Extended AFlow.}
AFlow~\cite{aflow} is a state-of-the-art agentic workflow framework that searches over code-represented workflows using MCTS with validation-set execution feedback. We extend it with specialized-model nodes and a cost/latency-aware objective, and follow~\cite{aflow} by running 20 MCTS iterations with \GPTFOUR{} as the optimizer and the same validation set size as \OURSYSTEM{}.

\para{\revision{Extended DyFlow.}}
\revision{DyFlow~\cite{dyflow} is an agentic workflow framework that iteratively designs workflows stage-by-stage with an LLM guided by validation-set feedback. We extend it with the same specialized-model nodes and cost/latency-aware objective, using \GPTFOUR{} as the driver and the same validation set size as \OURSYSTEM{}.}

\subsubsection{\revision{Validation Set Setup}} \label{sec:val-setup}

\revision{For fair comparison, all systems that consume a validation set---\OURSYSTEM{}, LOTUS, Palimpzest, DocETL, AFlow, and DyFlow---operate without human ground-truth labels: each uses an oracle LLM for proxy signals on unlabeled tuples.}

\begin{figure*}[t]
\centering
\includegraphics[width=\linewidth]{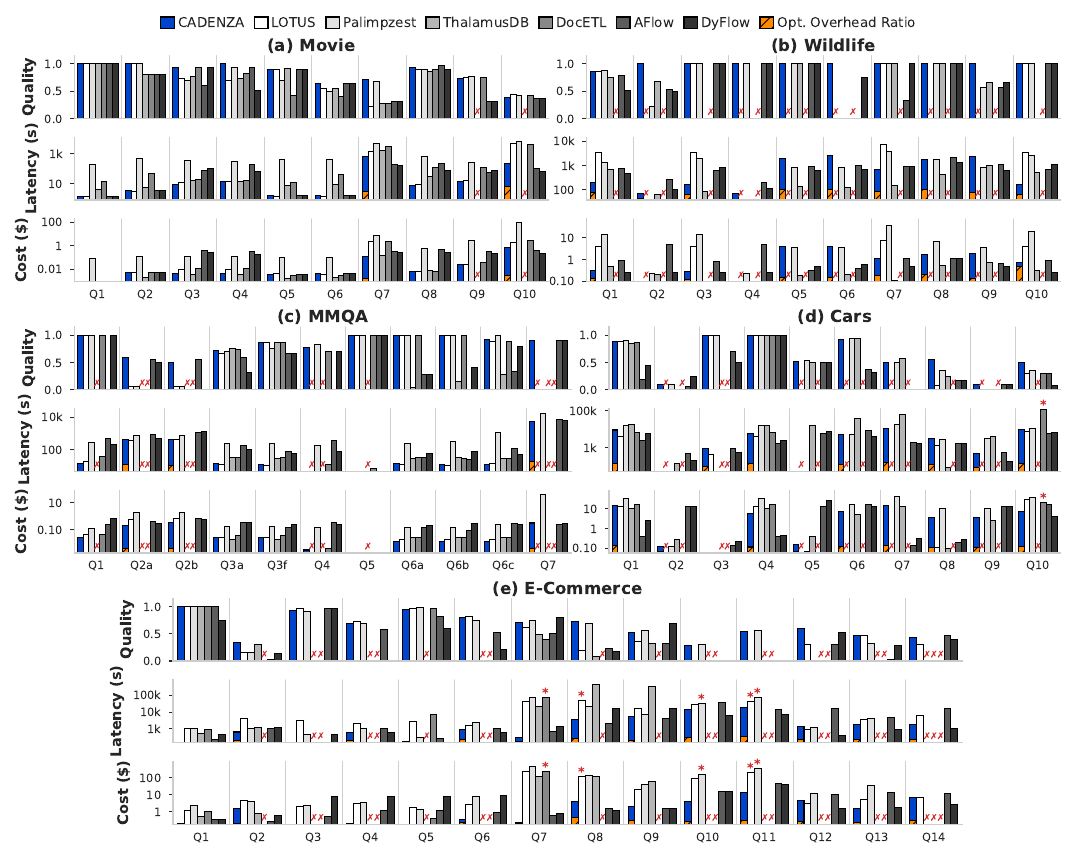}
\vspace{-8mm}
\caption{\revision{End-to-end per-query Q/L/C across systems for each scenario.}}
\label{fig:qlc_summary}
\vspace{-3mm}
\end{figure*}

\subsection{End-to-End Query Performance} \label{sec:endtoendperf}
Table~\ref{tab:semantic_op_benchmark_summary} summarizes \OURSYSTEM{}’s average relative performance across scenarios. Figure~\ref{fig:qlc_summary} visualizes per-query Q/L/C across systems. (\texttimes) denotes failures.
(*) marks latency/cost extrapolated from timed-out runs by rerunning at a smaller scale (E-Comm, SF500) and scaling by the input-cardinality ratio (quality from the smaller run). Hatched bar segments indicate the optimization overhead fraction of the total, including optimization and execution. Overall, \OURSYSTEM{} achieves consistently strong quality--latency--cost trade-offs---even on multi-operator queries (e.g., Q7 in Wildlife, Q10--12 in E-Comm).

\para{Movie (Figure~\ref{fig:qlc_summary}(a)).} On scenario-level averages, \OURSYSTEM{} improves Q by +0.15 (vs.\ DyFlow), L 21.0$\times$ / C 120.1$\times$ (vs.\ Palimpzest). For Q5, Q6, and Q7, \OURSYSTEM{} selects an \textsf{OpTextSenti}-based plan that requires far fewer LLM calls, resulting in low latency and cost. In contrast, LOTUS and DocETL use embedding similarity as a proxy, which is ill-suited for sentiment-based joins and leads to a notable drop in quality. Palimpzest executes LLM-based plans under a cost constraint by cascading from cheaper to more expensive models, improving quality over LOTUS but at a substantially higher cost. Palimpzest is slow and costly on Q8 and Q10 because it selects a Mixture-of-Agents~\cite{moa} plan, yet achieves only similar quality. \revision{AFlow and DyFlow remain competitive on easier cases but plateau on harder sentiment joins (e.g., Q7 F1 $=$ 0.31); their search procedures tend to settle on parameter tweaks rather than discover better structures within the budget, yielding high cross-query variance.} ThalamusDB skips Q9, Q10 (no ranking support). As shown in Table~\ref{tab:opt_overhead}, \OURSYSTEM{} incurs about 24\% latency and 17\% cost overhead, amortizing to ${\sim}9\%$ and ${\sim}6\%$.

\para{Wildlife (Figure~\ref{fig:qlc_summary}(b)).} \label{sec:wildlife}
On scenario-level averages, \OURSYSTEM{} improves Q by +0.48 (vs.\ ThalamusDB), L 20.8$\times$ / C 8.1$\times$ (vs.\ LOTUS).
LOTUS’s latency is dominated by upfront prompt materialization: it eagerly constructs per-row prompts for the entire input, which can take nearly 30 minutes.
Palimpzest avoids this upfront cost, but it often executes semantic filters with \GPTFOUR{}, making inference expensive; moreover, its optimization alone can cost over \$1 while offering little quality improvement over LOTUS.
In contrast, \OURSYSTEM{} selects plans with \textsf{OpImgCls} and tunes it to process data mostly by specialized backends, avoiding expensive LLM invocations and yielding faster, cheaper execution.
AFlow shows high quality variance (Q7), and MCTS overhead increases latency and cost overall.
As shown in Table~\ref{tab:opt_overhead}, \OURSYSTEM{} adds ${\sim}24\%$ latency and ${\sim}20\%$ cost overhead, amortizing to ${\sim}9\%$ and ${\sim}8\%$.
In Q3 and Q10, ThalamusDB is fast but inaccurate: its early stopping checks whether the answer is stable under both extreme completions (\texttt{default=0/1}), which for aggregation/ranking only indicates stability at the extremes and misses true-label concentrations that flip the top-1.

\para{\revision{MMQA (Figure~\ref{fig:qlc_summary}(c)).}}
\revision{Most queries operate on inputs small enough to fall under \OURSYSTEM{}'s bypass threshold. On the image--text logo joins (Q2a, Q2b), LOTUS, Palimpzest, and DyFlow apply a brute-force LLM oracle to every (image, text) pair, mistaking visually similar but distinct logos for matches and collapsing F1 to roughly 0.05--0.07. \OURSYSTEM{} instead chooses an \textsf{OpImgCls}-based join that scores pairs from visual representations and only escalates uncertain cases to an LLM verifier, lifting F1 to 0.59\,/\,0.50, on par with AFlow's specialized-model plan but without its MCTS overhead. On the large-scale join (Q7), LOTUS, ThalamusDB, and DocETL fail to complete, and Palimpzest's cost-driven cascade still enumerates every pair, while \OURSYSTEM{}, AFlow, and DyFlow all reach F1 0.91 by pruning pairs with cheap symbolic/specialized cues before any LLM call. ThalamusDB and DocETL lack operators required by several queries (map and image input queries, respectively).}

\para{\revision{Cars (Figure~\ref{fig:qlc_summary}(d)).}} \revision{Cars is the most modality-rich scenario---each car may carry text, image, and audio modalities on top of structured fields (transmission, crash, fuel\_type)---so the right plan typically gates an expensive modality operator behind a cheap structured filter. 
\OURSYSTEM{} picks exactly such plans: Q1 (``cars in a crash'') answers via the structured \texttt{crash} field alone (F1 0.89), and Q3 (``manual transmission, damaged in images'') prunes by transmission before running \textsf{OpImgCls} on survivors (F1 1.00). AFlow and DyFlow stick with image-classifier plans and miss the cheap structured 
filters, degrading to F1 0.19\,/\,0.44 on Q1 and 0.70\,/\,0.50 on Q3 
at no cost benefit; Q6 and Q7 show the same single-modality over-commit. Palimpzest cascades over the whole input rather than after a structured 
prefilter, spending \$15--\$33 on multi-modal queries to reach the F1 \OURSYSTEM{} hits at \$6--\$14. Q8 (``punctures and paint scratches on images'') is the clearest \OURSYSTEM{} win: every baseline conflates the two damage types under a single LLM oracle (F1 0.08--0.35; LOTUS asks the LLM to verify both per image and collapses to F1 0.08), while \OURSYSTEM{} stages a coarse \textsf{OpImgCls} scan followed by a selective LLM verifier on positives only, lifting F1 to 0.56. LOTUS, 
DocETL, and ThalamusDB cover only subsets of queries.}

\para{E-Commerce (Figure~\ref{fig:qlc_summary}(e)).} On scenario-level averages, \OURSYSTEM{} improves Q by +0.24 (vs.\ DyFlow), L 165.7$\times$ / C 310.3$\times$ (vs.\ DocETL). For Q1 (Reebok backpack filtering), we answer via simple keyword filtering over product name/description even without task-specific operators; for Q4 we use \textsf{OpImgCls} mainly with CLIP \cite{CLIP}. For Q3 and Q5 (text brand/category extraction), we use \textsf{OpTxtQA} and process it with lightweight keyword-based and BART~\cite{BART} zero-shot-based implementations for most inputs, routing only cue-missing or hard cases to an LLM backend. For Q6 (image category extraction), we use \textsf{OpImgCls} with cropping-based decomposition. Across these queries, the cheaper backends handle up to 91\% of inputs. For join-heavy queries (Q7--Q9), we adopt a similar pair-level strategy: prune pairs using cheap brand/category/color cues and then apply selective LLM filtering to avoid quadratic blowups; for Q8, we instead rely on \textsf{OpTextPairScore} paired with \textsf{OpImgCap}. In contrast, baselines are overly reliant on LLMs or ill-suited proxies (e.g., embedding similarity), and LOTUS further loses quality on Q8 due to lack of multi-key joins. Q8 is particularly challenging due to its large join input (28 texts and 4{,}993 images), causing most baselines to exceed the 24-hour timeout. For Palimpzest, we disabled Abacus on join-heavy queries because the optimization input exceeded the embedding model’s maximum sequence length. AFlow also uses specialized models (e.g., CLIP and BART), but it largely sticks to its initial plan and makes only minor tweaks, so quality improves little; meanwhile, MCTS overhead makes it slower and more expensive. ThalamusDB lacks a map and thus skips Q3--Q6.

\para{\revision{Takeaway.}} \revision{Current SQPEs do not expose a semantic operator instance's intermediate task outputs as a physical optimization object, and often over-spend on easy inputs.}

\subsection{Sensitivity Analysis} \label{sec:sensitivity}

To study sensitivity, we run analyses on the Movie scenario and report end-to-end performance as the geometric mean over queries where optimization is not skipped.

\subsubsection{Impact of Utility Weights} 

\begin{table}[ht]
\vspace{-4mm}
\centering
\small
\setlength{\tabcolsep}{7pt}
\renewcommand{\arraystretch}{1.10}
\caption{Sensitivity to utility weights.}
\vspace{-3mm}
\label{tab:sensitivity_weights}
\begin{tabular}{lccc}
\toprule
\textbf{$(w_{\text{q}}, w_{\text{l}}, w_{\text{c}})$} &
\textbf{Q}$\uparrow$ & \textbf{L (s)}$\downarrow$ & \textbf{C (\$)}$\downarrow$ \\
\midrule
$(1.0,\ 0.1,\ 0.1)$ & 0.78 & 4251 & 4.82 \\
$(0.1,\ 1.0,\ 1.0)$ & 0.41 &  104 & 0.08 \\
\textbf{$(1.0,\ 0.4,\ 0.3)$} & 0.65 & 355 & 0.30 \\
$(1.0,\ 0.9,\ 0.1)$ & 0.52 & 143 & 0.09 \\
$(1.0,\ 0.1,\ 0.9)$ & 0.59 & 181 & 0.09 \\
\bottomrule
\end{tabular}
\vspace{-2mm}
\end{table}

Table~\ref{tab:sensitivity_weights} reports sensitivity to the utility weights.
Quality-centric weights $(1.0, 0.1, 0.1)$ favor LLM-heavy plans, achieving the highest quality but also the largest latency and cost; BO can also slow down since many candidates remain LLM-intensive.
Latency/cost-centric weights $(0.1, 1.0, 1.0)$ largely avoids LLM calls and instead uses specialized or symbolic backends, reducing latency and cost at the expense of quality; the remaining cost mainly comes from early BO trials that still explore some LLM-including candidates.
For intermediate settings with $w_q{=}1.0$, increasing the latency weight (and lowering the cost weight) favors specialized/symbolic choices, yielding lower latency and low cost but reduced quality.
Shifting weight from latency to cost (from $(1.0, 0.9, 0.1)$ to $(1.0, 0.1, 0.9)$) keeps cost low but increases latency and quality, suggesting more reliance on specialized backends than symbolic ones while avoiding expensive LLM calls.
We use $(1.0,0.4,0.3)$ by default as a stable middle ground.

\subsubsection{Impact of BO Trials} 

\begin{figure}[ht]
    \vspace{-5mm}
    \centering
    \includegraphics[width=0.7\linewidth]{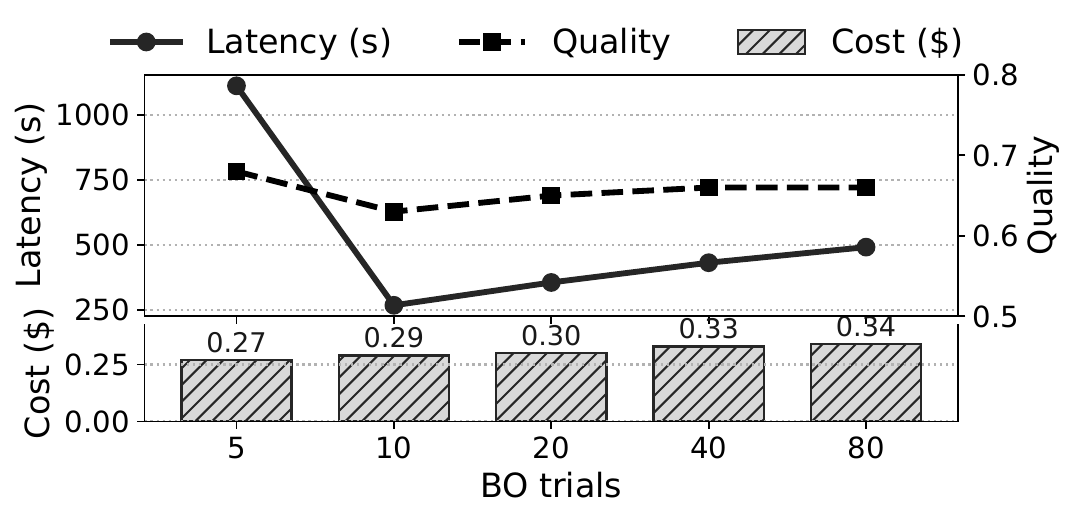}
    \vspace{-4mm}
    \caption{Sensitivity to Bayesian optimization trials.}
    \label{fig:bp_plot}
    \vspace{-4mm}
\end{figure}

Figure~\ref{fig:bp_plot} reports sensitivity to the BO trial budget. With our warm-start initialization, BO shows limited randomness, but with only 5 trials it may select a plan before revealing each candidate’s full potential. As a result, it can choose a plan using \textsf{OpTxtCls}, which yields slightly higher quality (e.g., 0.68) but higher latency due to a larger model, while the \textsf{OpTxtSenti}-based plan is not tuned enough to expose its latency advantage. With 10 trials, BO typically uncovers the latency potential of \textsf{OpTxtSenti}-based plan, leading to a large latency drop and making it the preferred choice despite a small quality loss. With 20 trials, additional tuning slightly improves quality (e.g., 0.63$\rightarrow$0.65) with router threshold tunings at the cost of extra tuning-time latency and cost, and beyond that further trials mainly increase tuning cost (dominated by LLM calls) without materially improving the final plan. In practice, the BO trial budget should depend on user intent and the size of the parameter space, which we leave for future work.

\para{\revision{Takeaway.}} \revision{Plan selection responds predictably to the utility weights, so users can dial the Q/L/C trade-off without per-query tuning, and Bayesian optimization converges within a moderate trial budget.}

\subsection{Ablation Study} \label{sec:ablation}

\revision{Table~\ref{tab:ablation_unified} reports a unified ablation that mixes additive build-up of the logical planning (rows 1--4, full physical) with leave-one-out on physical planning from Full \OURSYSTEM{} (rows 5--10). The ablation runs on SemBench together with BioDEX (end-to-end BioDEX results in Table~\ref{tab:appx:biodex}, Appendix~\ref{sec:appx:perscenario}); queries handled by \OURSYSTEM{}'s adaptive bypass are excluded from the aggregation. \textbf{BE} (Backends) $\in$ \{All, SYM, SP, GP, COM\}: All $=$ full pool, SYM $=$ symbolic only, SP $=$ specialized only, GP $=$ general-purpose LLM only, COM $=$ composite cascade. Router operates only over the All pool.}

\begin{table}[ht]
\vspace{-2mm}
\centering
\small
\setlength{\tabcolsep}{4pt}
\renewcommand{\arraystretch}{1.1}
\caption{\revision{Ablation study. Row 4 is Full \OURSYSTEM{}.}}
\vspace{-3mm}
\label{tab:ablation_unified}
\begin{tabular}{c c c c c c ccc}
\toprule
\textbf{\#} & \textbf{\#Seed} & \textbf{Expl} & \textbf{BO} & \textbf{Router} & \textbf{BE} & \textbf{Q}$\uparrow$ & \textbf{L (s)}$\downarrow$ & \textbf{C (\$)}$\downarrow$ \\
\midrule
1  & ---  & ---        & ---        & ---        & GP  & 0.64 & 9{,}086  & 29.2 \\
2  & 1    & $\times$   & \checkmark & \checkmark & All & 0.48 & 5{,}164  & 5.7  \\
3  & 3    & $\times$   & \checkmark & \checkmark & All & 0.67 & 2{,}743  & 2.7  \\
4  & 3    & \checkmark & \checkmark & \checkmark & All & 0.71 & 2{,}975  & 3.0 \\
\midrule
5  & 3    & \checkmark & $\times$   & \checkmark & All & 0.62 & 5{,}847  & 7.9  \\
6  & 3    & \checkmark & \checkmark & $\times$   & All & 0.63 & 6{,}896  & 16.6 \\
7  & 3    & \checkmark & \checkmark & ---        & SYM & 0.30 & 237      & 0.04 \\
8  & 3    & \checkmark & \checkmark & ---        & SP  & 0.50 & 1{,}448  & 0.21 \\
9  & 3    & \checkmark & \checkmark & ---        & GP  & 0.65 & 7{,}800  & 23.5 \\
10 & 3    & \checkmark & \checkmark & ---        & COM & 0.55 & 7{,}318  & 7.8  \\
\bottomrule
\end{tabular}
\vspace{-2mm}
\end{table}

\revision{The logical build-up (rows 1--3) carries most of the Q gain. At seed=1 (row 2), Q drops because a single plan often turns out structurally weak---operators may be miswired or backends mismatched---and full physical cannot rescue a broken plan. Multi-seed (row 3) is the dominant single jump: candidate diversity gives BO/Router something to exploit. Adding \textsc{Expl} (row 4) further raises Q on queries needing operator decomposition or modality switching, though the gain is amortized by queries already well-served.}

\revision{On the physical side, removing BO (row 5) is the most damaging: with the threshold tuner gone, Router defaults to erratic routing---easy cases escalate to GP and hard ones go to SYM. Removing Router (row 6) primarily inflates cost: BO can still pick the best backend per operator, but without per-tuple dispatch every input pays the same expensive tariff. Single-backend variants (rows 7--10) never recover Full's trade-off: GP reaches comparable Q at $8\times$ the cost; SYM is essentially free but only covers cue-rich queries.}

\para{\revision{Takeaway.}} \revision{Logical planning expands the achievable trade-off space; physical optimization---routing, backend-family choice, and BO-based tuning---determines which trade-off is realized.}

\section{Related Work} \label{sec:related_work} 

\para{Semantic Operator Optimization.}
LOTUS~\cite{LOTUS} approximates a reference LLM-based algorithm using proxy models. \revision{Palimpzest~\cite{palimpzest} integrates Abacus~\cite{abacus}, a cost-based optimizer over a user-specified program whose semantic operators are each bound to a single logical operator, searching over transformations and implementations for that fixed structure.} 
ThalamusDB~\cite{thalamusdb} targets AQP via partial results and early termination, again on a single logical operator binding. 
AOP~\cite{AOP} and Unify~\cite{unify} pre-register both general-purpose and symbolic backend implementations, using symbolic when applicable. 
\revision{DocETL~\cite{DOCETL} performs pipeline-level rewrites over document-processing operators (map, filter, reduce, split, gather, resolve, equijoin); the distinction is not whether a system can decompose, but what the decomposition exposes to the optimizer. DocETL does not expose decomposed task outputs as a relational physical optimization object, and its optimizer targets quality only.}
\revision{\OURSYSTEM{} instead compiles a semantic operator instance embedded in a relational query into a typed \TxRA{} DAG, exposing intermediate task outputs as relational attributes that physical search can filter, threshold, reorder, bind per-task to symbolic, specialized, general-purpose, or composite backends, and jointly tune for Q/L/C.}

\para{LLM-based Query Planning.}
LLMs are widely used to construct multi-step plans, either by decomposing tasks into sub-steps (e.g., DECOMP~\cite{decomposed-prompting}) or by interleaving reasoning with tool use (e.g., ReAct~\cite{react}, ADAPT~\cite{adapt}, Dart-LLM~\cite{dart-llm}). These approaches primarily target feasibility and typically lack transformation rules or explicit quality--cost optimization. Related systems refine programs or pipelines to improve end-task performance (e.g., MLE-STAR~\cite{mle-star}, Data-Interpreter~\cite{data-interpreter}, ADAS~\cite{adas}). Closer to optimization, Archon~\cite{archon} applies Bayesian optimization, AFlow~\cite{aflow} uses Monte Carlo Tree Search, \revision{and DyFlow~\cite{dyflow} uses an LLM-driven dynamic redesign loop}, but their search spaces are largely limited to \revision{code-represented LLM-call workflows} and do not incorporate relational-style optimization or heterogeneous backends. In contrast, \OURSYSTEM{} optimizes over heterogeneous backends in an extended relational algebra under an explicit quality--latency--cost objective.

\para{UDF Optimization.}
Several systems extend SQL engines with ML/LLM capabilities by treating semantic processing as expensive UDFs that encapsulate model calls, e.g., Caesura~\cite{caesura}, Amazon Redshift ML~\cite{aws-redshift-ml}, BigQuery~\cite{bigquery}, EVA~\cite{eva}, VOCAL-UDF~\cite{vocal-udf}, and Aero~\cite{aero}. They typically optimize a \emph{fixed} logical plan and treat UDFs as black boxes, focusing on efficient execution; for example, Aero~\cite{aero} uses adaptive execution to choose evaluation orders for black-box UDFs. In contrast, \OURSYSTEM{} exposes semantic processing via task-extended relational algebra, enabling intent-dependent decomposition and optimization rather than a monolithic UDF. Prior work on opening black-box UDFs (e.g., SQL inlining~\cite{prism, optudf}, control-flow graph decomposition~\cite{graceful}, and reorderability analysis~\cite{opening}) is not designed for model-backed, task-specific operators and does not address intent-dependent plan synthesis or multi-objective tuning.

\section{Conclusion} \label{sec:conclusion}

\revision{We presented \OURSYSTEM{}, a semantic operator optimizer that compiles each instance into a typed \TxRA{} DAG, exposing intermediate task outputs as relational attributes for routing across heterogeneous backends and joint tuning via Bayesian optimization. \OURSYSTEM{} achieves up to +0.49 quality and 165.7$\times$/310.3$\times$ latency/cost improvements on SemBench scenarios.}

% \begin{acks}

% \end{acks}

\bibliographystyle{ACM-Reference-Format}
\bibliography{bib/refs}

\fi

\ifnum\compilemode>0

\newpage
\appendix

\ifnum\compilemode=2
    \setcounter{page}{1}
\fi

\clearpage
\onecolumn

\section{Scaling with Data Size} \label{sec:scalability}

To study scaling, we vary data size in the Movie with all other settings fixed, and report end-to-end metrics as the geometric mean over queries where optimization is not skipped. In Figure~\ref{fig:scalability_plot}, optimization dominates at small scales (SF=10K: $\sim$52\% of total latency/cost) but drops for SF$\ge$20K, \jhha{since it runs on a fixed-size validation set under a fixed BO budget and is thus largely scale-independent}. In contrast, latency/cost increase with data size and can be superlinear for join queries due to growing intermediate cardinalities. This implies that the optimization overhead is largely scale-independent and amortizes as data grows; plan caching and adaptive bypass further reduce overhead for repeated intents and small inputs.

\begin{figure}[ht]
    \vspace{-3mm}
    \centering
    \includegraphics[width=0.5\linewidth]{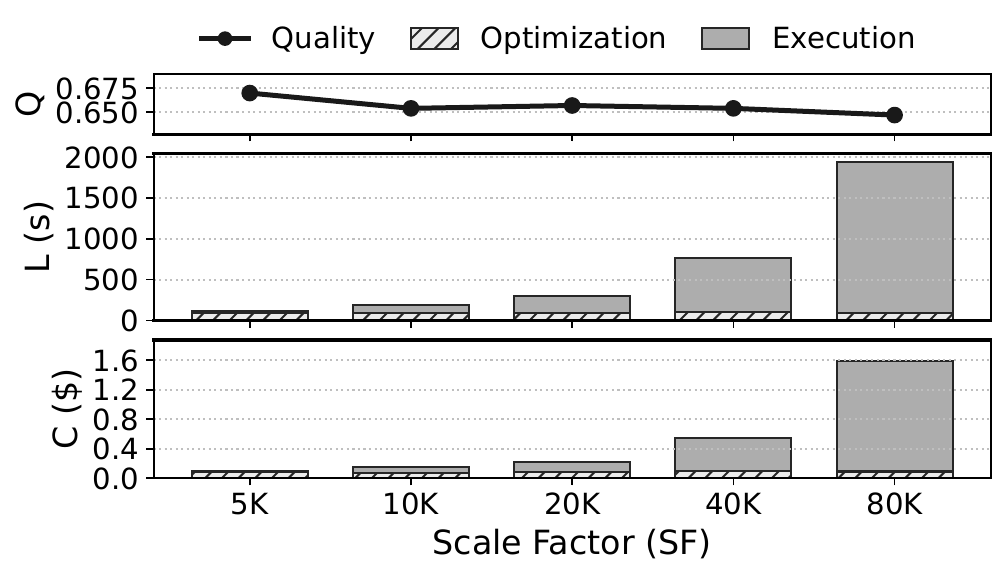}
    \vspace{-4mm}
    \caption{End-to-end scaling with data size.}
    \label{fig:scalability_plot}
    \vspace{-6mm}
\end{figure}

\section{\revision{Per-Scenario and BioDEX End-to-End Results}} \label{sec:appx:perscenario}

\revision{Tables~\ref{tab:appx:movie}--\ref{tab:appx:ecomm} report the per-scenario raw values summarized in Table~\ref{tab:semantic_op_benchmark_summary}. Each cell is formatted as \emph{Competitor}\,/\,\OURSYSTEM{}, with \OURSYSTEM{}'s statistics computed over the intersection of queries handled by that competitor for fair comparison (\OURSYSTEM{} itself is a standalone optimizer/executor; see \S\ref{sec:impl:stack}). Bold per cell marks the better value (higher Quality, lower Latency/Cost); ties (within 0.005 quality and 0.5 s/\$) are unmarked.}

\revision{Table~\ref{tab:appx:biodex} additionally reports results on the BioDEX benchmark~\cite{biodex}, an extreme multi-label classification task that identifies adverse drug reactions from medical articles, modeled as a \join{} between articles and reactions~\cite{LOTUS}. We use a sampled subset of 1{,}000 articles and 100 reactions, with each article constructed from its title, abstract, and relevant sections; quality is measured by Rank Precision@K (RP@K) with $K=5$~\cite{LOTUS, abacus}. Cell format and bolding follow the per-scenario tables.}

\noindent
\begin{minipage}[t]{0.49\linewidth}
\begin{table}[H]
\caption{\revision{Movie scenario.}}
\label{tab:appx:movie}
\centering
\setlength{\tabcolsep}{4pt}
\renewcommand{\arraystretch}{1.05}
\begin{tabular}{lccc}
\toprule
\textbf{System} & \textbf{Q}$\uparrow$ & \textbf{L (s)}$\downarrow$ & \textbf{C (\$)}$\downarrow$ \\
\midrule
LOTUS              & 0.72\,/\,\textbf{0.82} & 17\,/\,\textbf{10}        & 0.02\,/\,\textbf{0.01} \\
Palimpzest         & 0.76\,/\,\textbf{0.82} & 636\,/\,\textbf{10}       & 0.53\,/\,\textbf{0.01} \\
ThalamusDB         & 0.74\,/\,\textbf{0.89} & 20\,/\,\textbf{7}         & \textbf{0.00}\,/\,0.01 \\
DocETL             & 0.67\,/\,\textbf{0.82} & 88\,/\,\textbf{10}        & 0.02\,/\,\textbf{0.01} \\
Extended AFlow     & 0.68\,/\,\textbf{0.82} & 24\,/\,\textbf{10}        & 0.05\,/\,\textbf{0.01} \\
Extended DyFlow    & 0.67\,/\,\textbf{0.82} & 19\,/\,\textbf{10}        & 0.04\,/\,\textbf{0.01} \\
\bottomrule
\end{tabular}
\end{table}
\end{minipage}\hfill
\begin{minipage}[t]{0.49\linewidth}
\begin{table}[H]
\caption{\revision{Wildlife scenario.}}
\label{tab:appx:wildlife}
\centering
\setlength{\tabcolsep}{4pt}
\renewcommand{\arraystretch}{1.05}
\begin{tabular}{lccc}
\toprule
\textbf{System} & \textbf{Q}$\uparrow$ & \textbf{L (s)}$\downarrow$ & \textbf{C (\$)}$\downarrow$ \\
\midrule
LOTUS              & 0.96\,/\,0.97          & 4{,}022\,/\,\textbf{156}  & 4.64\,/\,\textbf{0.51} \\
Palimpzest         & 0.77\,/\,\textbf{0.99} & 727\,/\,\textbf{326}      & 4.30\,/\,\textbf{0.62} \\
ThalamusDB         & 0.51\,/\,\textbf{0.99} & \textbf{154}\,/\,326      & \textbf{0.17}\,/\,0.62 \\
DocETL             & ---                    & ---                       & ---                     \\
Extended AFlow     & 0.72\,/\,\textbf{0.99} & 703\,/\,\textbf{326}      & 1.06\,/\,\textbf{0.62} \\
Extended DyFlow    & 0.84\,/\,\textbf{0.99} & 527\,/\,\textbf{326}      & \textbf{0.38}\,/\,0.62 \\
\bottomrule
\end{tabular}
\end{table}
\end{minipage}

\vspace{-2mm}
\noindent
\begin{minipage}[t]{0.49\linewidth}
\begin{table}[H]
\caption{\revision{MMQA scenario.}}
\label{tab:appx:mmqa}
\centering
\setlength{\tabcolsep}{4pt}
\renewcommand{\arraystretch}{1.05}
\begin{tabular}{lccc}
\toprule
\textbf{System} & \textbf{Q}$\uparrow$ & \textbf{L (s)}$\downarrow$ & \textbf{C (\$)}$\downarrow$ \\
\midrule
LOTUS              & 0.72\,/\,\textbf{0.84} & 22\,/\,\textbf{21}        & 0.04\,/\,\textbf{0.02} \\
Palimpzest         & 0.67\,/\,\textbf{0.84} & 348\,/\,\textbf{29}       & 0.28\,/\,\textbf{0.03} \\
ThalamusDB         & 0.41\,/\,\textbf{0.90} & 30\,/\,\textbf{13}        & 0.02\,/\,\textbf{0.02} \\
DocETL             & 0.90\,/\,\textbf{0.91} & 23\,/\,\textbf{8}         & 0.02\,/\,\textbf{0.01} \\
Extended AFlow     & 0.45\,/\,\textbf{0.84} & 169\,/\,\textbf{29}       & 0.15\,/\,\textbf{0.03} \\
Extended DyFlow    & 0.60\,/\,\textbf{0.84} & 126\,/\,\textbf{29}       & 0.19\,/\,\textbf{0.03} \\
\bottomrule
\end{tabular}
\end{table}
\end{minipage}\hfill
\begin{minipage}[t]{0.49\linewidth}
\begin{table}[H]
\caption{\revision{Cars scenario.}}
\label{tab:appx:cars}
\centering
\setlength{\tabcolsep}{4pt}
\renewcommand{\arraystretch}{1.05}
\begin{tabular}{lccc}
\toprule
\textbf{System} & \textbf{Q}$\uparrow$ & \textbf{L (s)}$\downarrow$ & \textbf{C (\$)}$\downarrow$ \\
\midrule
LOTUS              & 0.65\,/\,\textbf{0.79} & \textbf{2{,}402}\,/\,3{,}467   & \textbf{1.23}\,/\,1.63 \\
Palimpzest         & 0.57\,/\,\textbf{0.61} & 1{,}547\,/\,\textbf{1{,}209}  & 3.40\,/\,\textbf{1.57} \\
ThalamusDB         & 0.51\,/\,\textbf{0.57} & 5{,}235\,/\,\textbf{1{,}019}  & \textbf{1.93}\,/\,2.66 \\
DocETL             & 0.72\,/\,\textbf{0.80} & 17{,}217\,/\,\textbf{6{,}415} & 17.15\,/\,\textbf{8.50} \\
Extended AFlow     & 0.34\,/\,\textbf{0.61} & 2{,}030\,/\,\textbf{1{,}209}  & 1.76\,/\,\textbf{1.57} \\
Extended DyFlow    & 0.34\,/\,\textbf{0.61} & 1{,}730\,/\,\textbf{1{,}209}  & 2.34\,/\,\textbf{1.57} \\
\bottomrule
\end{tabular}
\end{table}
\end{minipage}

\vspace{-2mm}
\noindent
\begin{minipage}[t]{0.49\linewidth}
\begin{table}[H]
\caption{\revision{E-Commerce scenario.}}
\label{tab:appx:ecomm}
\centering
\setlength{\tabcolsep}{4pt}
\renewcommand{\arraystretch}{1.05}
\begin{tabular}{lccc}
\toprule
\textbf{System} & \textbf{Q}$\uparrow$ & \textbf{L (s)}$\downarrow$ & \textbf{C (\$)}$\downarrow$ \\
\midrule
LOTUS              & 0.49\,/\,\textbf{0.64} & 6{,}164\,/\,\textbf{829}   & 10.18\,/\,\textbf{1.05} \\
Palimpzest         & 0.59\,/\,\textbf{0.65} & 3{,}940\,/\,\textbf{792}   & 18.44\,/\,\textbf{0.91} \\
ThalamusDB         & 0.44\,/\,\textbf{0.66} & 17{,}767\,/\,\textbf{555}  & 12.42\,/\,\textbf{0.91} \\
DocETL             & 0.83\,/\,\textbf{0.89} & 3{,}803\,/\,\textbf{85}    & 0.55\,/\,\textbf{0.17} \\
Extended AFlow     & 0.41\,/\,\textbf{0.64} & 2{,}125\,/\,\textbf{829}   & 2.28\,/\,\textbf{1.05} \\
Extended DyFlow    & 0.39\,/\,\textbf{0.64} & 1{,}335\,/\,\textbf{829}   & 2.98\,/\,\textbf{1.05} \\
\bottomrule
\end{tabular}
\end{table}
\end{minipage}\hfill
\begin{minipage}[t]{0.49\linewidth}
\begin{table}[H]
\caption{\revision{BioDEX (Q $=$ RP@5).}}
\label{tab:appx:biodex}
\centering
\setlength{\tabcolsep}{4pt}
\renewcommand{\arraystretch}{1.05}
\begin{tabular}{lccc}
\toprule
\textbf{System} & \textbf{Q}$\uparrow$ & \textbf{L (s)}$\downarrow$ & \textbf{C (\$)}$\downarrow$ \\
\midrule
LOTUS              & 0.16\,/\,\textbf{0.31}  & 4{,}111\,/\,3{,}925           & 61.74\,/\,\textbf{4.62} \\
Palimpzest         & 0.20\,/\,\textbf{0.31}  & 10{,}216\,/\,\textbf{3{,}925} & 84.45\,/\,\textbf{4.62} \\
ThalamusDB         & 0.03\,/\,\textbf{0.31}  & 5{,}948\,/\,\textbf{3{,}925}  & 5.82\,/\,\textbf{4.62} \\
DocETL             & 0.27\,/\,\textbf{0.31}  & 16{,}751\,/\,\textbf{3{,}925} & 23.83\,/\,\textbf{4.62} \\
Extended AFlow     & 0.17\,/\,\textbf{0.31}  & 4{,}256\,/\,\textbf{3{,}925}  & \textbf{0.43}\,/\,4.62 \\
Extended DyFlow    & 0.22\,/\,\textbf{0.31}  & \textbf{3{,}492}\,/\,3{,}925  & \textbf{0.72}\,/\,4.62 \\
\bottomrule
\end{tabular}
\end{table}
\end{minipage}

\section{\revision{Importance of Intent-specific Synthesis}} \label{sec:intent_specific_synthesis}

\revision{Tables~\ref{tab:appx:preset:movie}--\ref{tab:appx:preset:ecomm} compare \OURSYSTEM{} against \emph{Coarse}, a baseline that pre-declares one logical/physical plan per preference profile and routes each row between a strong LLM and a cheaper helper---an embedding cascade for filter/join, a nano LLM for map---at a fixed LLM share (90\%/50\%/10\% for Quality-first/Balanced/Efficiency-first); Coarse exposes the same preference interface as \OURSYSTEM{} but skips planning. Queries handled by \OURSYSTEM{}'s adaptive bypass are excluded; Q is an arithmetic mean across queries, while L and C are geometric means. Cells are \textsc{Coarse}\,/\,\OURSYSTEM{}; \textbf{bold} marks the better metric. \OURSYSTEM{} dominates Coarse on every metric in every (scenario, preference) cell. The gap traces to three missing ingredients: \emph{(i) no tuned routing}---Coarse applies a fixed plan per preference uniformly across queries (even LOTUS tunes a per-query cascade threshold), so easy queries pay the LLM tariff while hard ones are starved of evidence; \emph{(ii) no operator decomposition}---the single template$\to$operator binding cannot rewrite into intent-specific DAGs (e.g., a structured prefilter before image inference, or a coarse classifier followed by an LLM verifier); \emph{(iii) no backend diversity}---pooling choices into one strong LLM and one cheap helper forgoes the symbolic/specialized/general/composite mix that \OURSYSTEM{} dispatches per tuple. Intent-specific synthesis, not the preference interface, drives the gain.}

\noindent
\begin{minipage}[t]{0.49\linewidth}
\begin{table}[H]
\caption{\revision{Movie scenario.}}
\label{tab:appx:preset:movie}
\centering
\setlength{\tabcolsep}{4pt}
\renewcommand{\arraystretch}{1.05}
\begin{tabular}{lccc}
\toprule
\textbf{Preference} & \textbf{Q}$\uparrow$ & \textbf{L (s)}$\downarrow$ & \textbf{C (\$)}$\downarrow$ \\
\midrule
Quality-first    & 0.514 / \textbf{0.559} & 3{,}077 / \textbf{1{,}935} & \$3.29 / \textbf{\$0.85} \\
Balanced         & 0.425 / \textbf{0.544} & 3{,}594 / \textbf{266}     & \$2.01 / \textbf{\$0.30} \\
Efficiency-first & 0.201 / \textbf{0.447} & 1{,}743 / \textbf{198}     & \$0.72 / \textbf{\$0.15} \\
\bottomrule
\end{tabular}
\end{table}
\end{minipage}\hfill
\begin{minipage}[t]{0.49\linewidth}
\begin{table}[H]
\caption{\revision{Wildlife scenario.}}
\label{tab:appx:preset:wildlife}
\centering
\setlength{\tabcolsep}{4pt}
\renewcommand{\arraystretch}{1.05}
\begin{tabular}{lccc}
\toprule
\textbf{Preference} & \textbf{Q}$\uparrow$ & \textbf{L (s)}$\downarrow$ & \textbf{C (\$)}$\downarrow$ \\
\midrule
Quality-first    & 0.711 / \textbf{0.963} & 814 / \textbf{384} & \$1.85 / \textbf{\$0.77} \\
Balanced         & 0.617 / \textbf{0.986} & 554 / \textbf{326} & \$1.23 / \textbf{\$0.62} \\
Efficiency-first & 0.392 / \textbf{0.908} & 342 / \textbf{212} & \$0.80 / \textbf{\$0.34} \\
\bottomrule
\end{tabular}
\end{table}
\end{minipage}

\vspace{-2mm}
\noindent
\begin{minipage}[t]{0.49\linewidth}
\begin{table}[H]
\caption{\revision{MMQA scenario.}}
\label{tab:appx:preset:mmqa}
\centering
\setlength{\tabcolsep}{4pt}
\renewcommand{\arraystretch}{1.05}
\begin{tabular}{lccc}
\toprule
\textbf{Preference} & \textbf{Q}$\uparrow$ & \textbf{L (s)}$\downarrow$ & \textbf{C (\$)}$\downarrow$ \\
\midrule
Quality-first    & 0.206 / \textbf{0.692} & 1{,}950 / \textbf{920} & \$0.81 / \textbf{\$0.34} \\
Balanced         & 0.170 / \textbf{0.666} & 1{,}326 / \textbf{780} & \$0.54 / \textbf{\$0.27} \\
Efficiency-first & 0.111 / \textbf{0.613} & 819 / \textbf{507}     & \$0.35 / \textbf{\$0.15} \\
\bottomrule
\end{tabular}
\end{table}
\end{minipage}\hfill
\begin{minipage}[t]{0.49\linewidth}
\begin{table}[H]
\caption{\revision{Cars scenario.}}
\label{tab:appx:preset:cars}
\centering
\setlength{\tabcolsep}{4pt}
\renewcommand{\arraystretch}{1.05}
\begin{tabular}{lccc}
\toprule
\textbf{Preference} & \textbf{Q}$\uparrow$ & \textbf{L (s)}$\downarrow$ & \textbf{C (\$)}$\downarrow$ \\
\midrule
Quality-first    & 0.445 / \textbf{0.619} & 3{,}023 / \textbf{1{,}427} & \$4.70 / \textbf{\$1.96} \\
Balanced         & 0.390 / \textbf{0.609} & 2{,}056 / \textbf{1{,}209} & \$3.13 / \textbf{\$1.57} \\
Efficiency-first & 0.244 / \textbf{0.560} & 1{,}270 / \textbf{786}     & \$2.04 / \textbf{\$0.86} \\
\bottomrule
\end{tabular}
\end{table}
\end{minipage}

\vspace{-2mm}
\noindent
\begin{minipage}[t]{0.49\linewidth}
\begin{table}[H]
\caption{\revision{E-Commerce scenario.}}
\label{tab:appx:preset:ecomm}
\centering
\setlength{\tabcolsep}{4pt}
\renewcommand{\arraystretch}{1.05}
\begin{tabular}{lccc}
\toprule
\textbf{Preference} & \textbf{Q}$\uparrow$ & \textbf{L (s)}$\downarrow$ & \textbf{C (\$)}$\downarrow$ \\
\midrule
Quality-first    & 0.504 / \textbf{0.658} & 2{,}072 / \textbf{978} & \$3.15 / \textbf{\$1.31} \\
Balanced         & 0.442 / \textbf{0.639} & 1{,}409 / \textbf{829} & \$2.10 / \textbf{\$1.05} \\
Efficiency-first & 0.270 / \textbf{0.587} & 870 / \textbf{539}     & \$1.36 / \textbf{\$0.58} \\
\bottomrule
\end{tabular}
\end{table}
\end{minipage}\hfill
\begin{minipage}[t]{0.49\linewidth}
\hfill
\end{minipage}

\section{\revision{Cross-System Utility-Weight Sweep}} \label{sec:appx:utility-sweep}

\revision{The end-to-end results in \S\ref{sec:endtoendperf} report Q, L, and C separately at the default weight only. We therefore evaluate every system under the same utility $U$ from \S\ref{sec:overview:moop} and sweep five weight settings across all SemBench scenarios, asking whether \OURSYSTEM{}'s advantage is robust across representative preferences.}

\para{\revision{Setup.}}
\revision{We evaluate all five SemBench scenarios (Movie, Wildlife, MMQA, Cars, E-Comm). Per scenario, queries handled by \OURSYSTEM{}'s adaptive bypass are excluded. For each (system, query, weight), we compute $U$ from \S\ref{sec:overview:moop} with $\hat{L}$ and $\hat{C}$ obtained by per-query min--max normalisation across all (system, weight) runs, then average within each scenario. Weights, in $(w_q, w_l, w_c)$ order, span balanced, quality-leaning, efficiency-leaning, equal, and quality-only regimes: $W_1{=}(1,0.4,0.3)$, $W_2{=}(1,0.1,0.1)$, $W_3{=}(0.3,1,1)$, $W_4{=}(1,1,1)$, $W_5{=}(1,0,0)$. \OURSYSTEM{}, Extended AFlow, and Extended DyFlow ingest the weights directly. Palimpzest's optimiser receives them via a custom \texttt{WeightedSum} policy. LOTUS and ThalamusDB lack explicit utility weights; we sweep their native quality knob (LOTUS: \texttt{recall\_target}; ThalamusDB: \texttt{max\_error}) and, per weight, retain the frontier point with the highest $U$. DocETL is excluded as a quality-only optimiser.}

\begin{figure}[H]
\centering
\includegraphics[width=\linewidth]{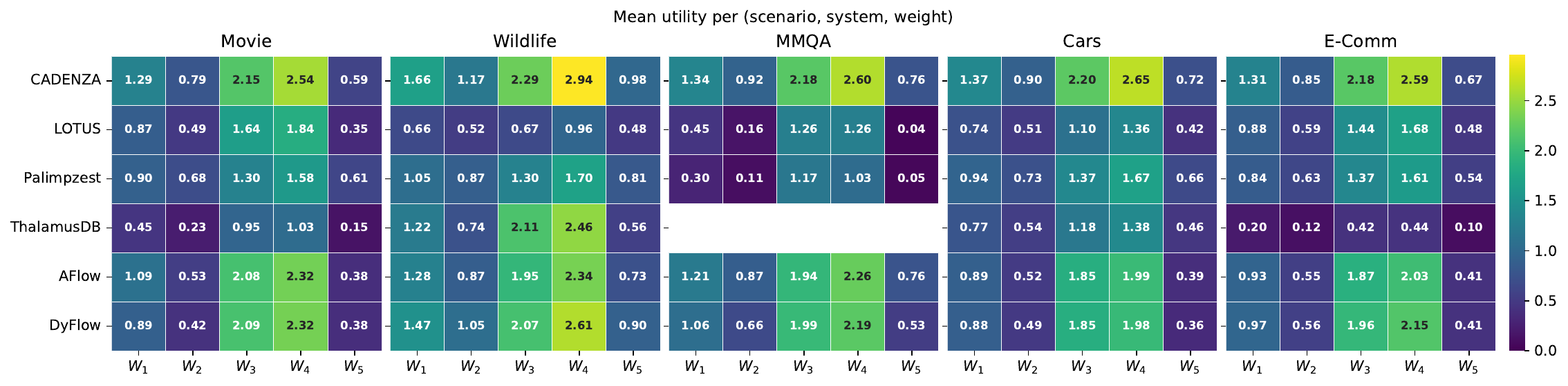}
\caption{\revision{Mean utility per (scenario, system, weight). Queries a system fails to run contribute zero to its mean (zero-fill). Higher is better.}}
\label{fig:utility-heatmap}
\vspace{-3mm}
\end{figure}

\vspace{-2mm}
\begin{table}[H]
\centering
\caption{\revision{Per-(system, weight) Quality, Latency, Cost---pooled across all (scenario, non-bypass query) tuples. Quality is the arithmetic mean of $\hat Q$ (F1 or $1\!-\!\mathrm{MAPE}/100$); Latency and Cost are geometric means.}}
\label{tab:utility-sweep}
\small
\setlength{\tabcolsep}{3pt}
\renewcommand{\arraystretch}{1.1}
\begin{tabular}{l ccc ccc ccc ccc ccc}
\toprule
& \multicolumn{3}{c}{$W_1$} & \multicolumn{3}{c}{$W_2$} & \multicolumn{3}{c}{$W_3$} & \multicolumn{3}{c}{$W_4$} & \multicolumn{3}{c}{$W_5$} \\
\cmidrule(lr){2-4} \cmidrule(lr){5-7} \cmidrule(lr){8-10} \cmidrule(lr){11-13} \cmidrule(lr){14-16}
\textbf{System} & Q & L & C & Q & L & C & Q & L & C & Q & L & C & Q & L & C \\
\midrule
\textbf{\OURSYSTEM{}}    & 0.73 & 9.9m  & \$0.58  & 0.75 & 11.8m & \$0.72  & 0.72 & 7.4m  & \$0.39 & 0.73 & 8.6m  & \$0.49  & 0.76 & 13.1m & \$0.81 \\
LOTUS                    & 0.53 & 1.1h  & \$4.70  & 0.54 & 1.1h  & \$4.98  & 0.50 & 57.1m & \$3.99 & 0.51 & 1.0h  & \$4.32  & 0.55 & 1.2h  & \$5.17 \\
Palimpzest               & 0.59 & 49.9m & \$13.28 & 0.61 & 52.5m & \$14.14 & 0.52 & 38.9m & \$9.63 & 0.56 & 42.9m & \$11.32 & 0.61 & 1.1h  & \$17.83 \\
ThalamusDB               & 0.50 & 42.1m & \$1.48  & 0.51 & 45.5m & \$1.63  & 0.48 & 31.6m & \$1.04 & 0.49 & 37.9m & \$1.30  & 0.51 & 48.4m & \$1.75 \\
Extended AFlow           & 0.47 & 26.7m & \$1.17  & 0.50 & 32.6m & \$1.48  & 0.42 & 18.6m & \$0.75 & 0.45 & 22.2m & \$0.96  & 0.51 & 37.9m & \$1.82 \\
Extended DyFlow          & 0.47 & 19.2m & \$1.20  & 0.50 & 25.2m & \$1.61  & 0.44 & 13.9m & \$0.80 & 0.47 & 16.5m & \$1.01  & 0.52 & 26.9m & \$1.79 \\
\bottomrule
\end{tabular}
\end{table}

\para{\revision{Result.}}
\revision{Figure~\ref{fig:utility-heatmap} shows that \OURSYSTEM{} produces the highest or tied-highest mean utility on every (scenario, weight) cell. The margin is largest on cost-aware settings ($W_3$, $W_4$), where Table~\ref{tab:utility-sweep} confirms that no baseline matches \OURSYSTEM{}'s pooled cost; on quality-leaning settings ($W_2$, $W_5$), \OURSYSTEM{} reaches higher Q at an order-of-magnitude lower C than the nearest competitor. Each baseline's limited weight-responsiveness stems from missing one of \OURSYSTEM{}'s three ingredients---operator catalog, BO-tuned plan parameters, and intent-aware plan synthesis. LOTUS only shifts its LLM-vs.-proxy cascade threshold (\texttt{recall\_target}). ThalamusDB only tunes its sampling budget (\texttt{max\_error}); the operator choice never changes. Palimpzest searches a richer plan space, but its candidate set is a small discrete pool of pre-computed physical plans with no continuous tuning. Extended AFlow and DyFlow search a workflow space, but their grammar is narrower than \OURSYSTEM{}'s catalog and their LLM-driven code generation can produce broken candidates. Each baseline therefore traverses only the subspace its mechanism exposes.}

\section{Pareto Frontier Approximation} \label{sec:pareto}

We evaluate whether \OURSYSTEM{} can recover a good approximate Pareto frontier over quality, latency, and cost. Figure~\ref{fig:pareto_hv} (top) shows the BO outcomes for one representative query (Movie, Q5): each point is one evaluated choice $\omega_t\in\Omega$ with its measured performance on the validation set, where the x-axis is normalized latency, the y-axis is quality, and color encodes normalized cost. Stars mark the final non-dominated Pareto set among all evaluations, and we label milestone trials (3, 5, 13, 21) where the frontier visibly expands.

\begin{figure}[ht]
    \centering
    \includegraphics[width=0.6\linewidth]{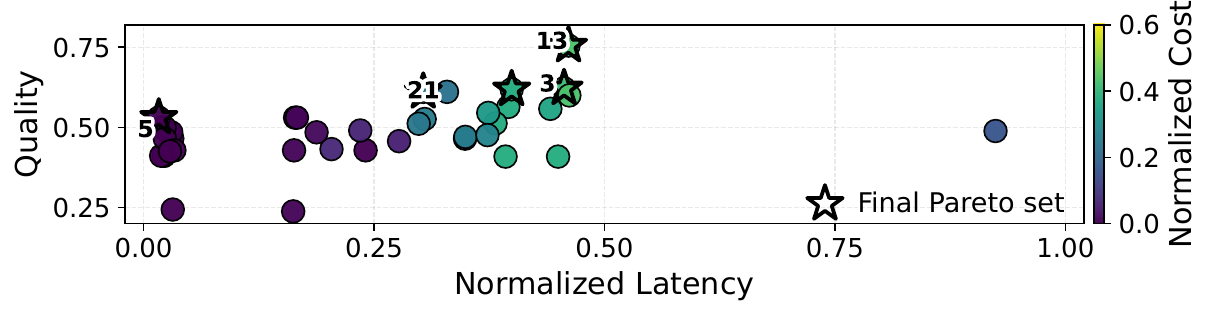}
    
    \includegraphics[width=0.6\linewidth]{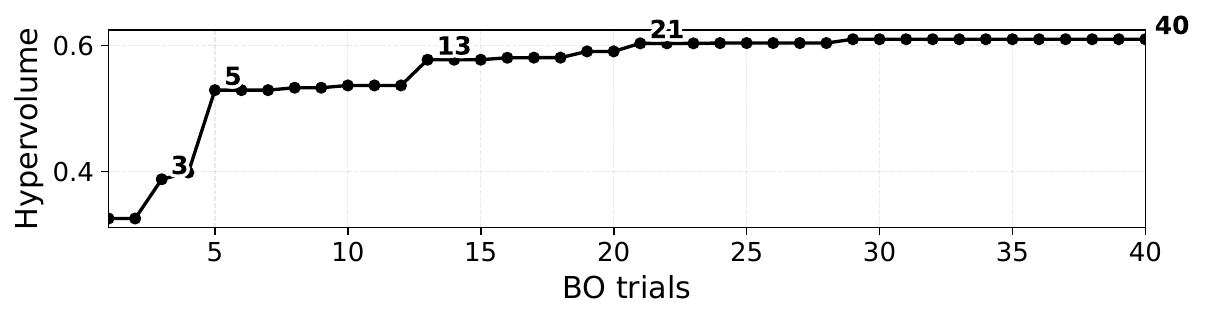}
    \caption{BO trials and hypervolume over the Pareto set.}
    \label{fig:pareto_hv}
\end{figure}

To quantify how the recovered frontier improves over BO, we track hypervolume after each trial.
Using the same normalization and maximization space as above (with the same $L_{\max}$ and $C_{\max}$),
let $B_{tune}$ be the total BO trial budget and let $\mathcal{F}_t$ denote the non-dominated points among the first $t$ trials ($t=1,\dots,B$).
With a dominated reference point $r=(0,0,0)$, we compute
\[
HV(\mathcal{F}_t; r)
= \lambda\!\left(\bigcup_{x\in\mathcal{F}_t}
[r_q,x_q]\times[r_\ell,x_\ell]\times[r_c,x_c]\right),
\]
where $x=(x_q,x_\ell,x_c)$ and $\lambda(\cdot)$ denotes 3D volume.
Figure~\ref{fig:pareto_hv} (bottom) plots $HV(\mathcal{F}_t; r)$ versus $t$; steps indicate trials that expand the frontier, while plateaus indicate no improvement.

\section{Plan Visualization} \label{sec:plan_visualization}

\begin{figure}[ht]
  \centering
  \includegraphics[width=\linewidth]{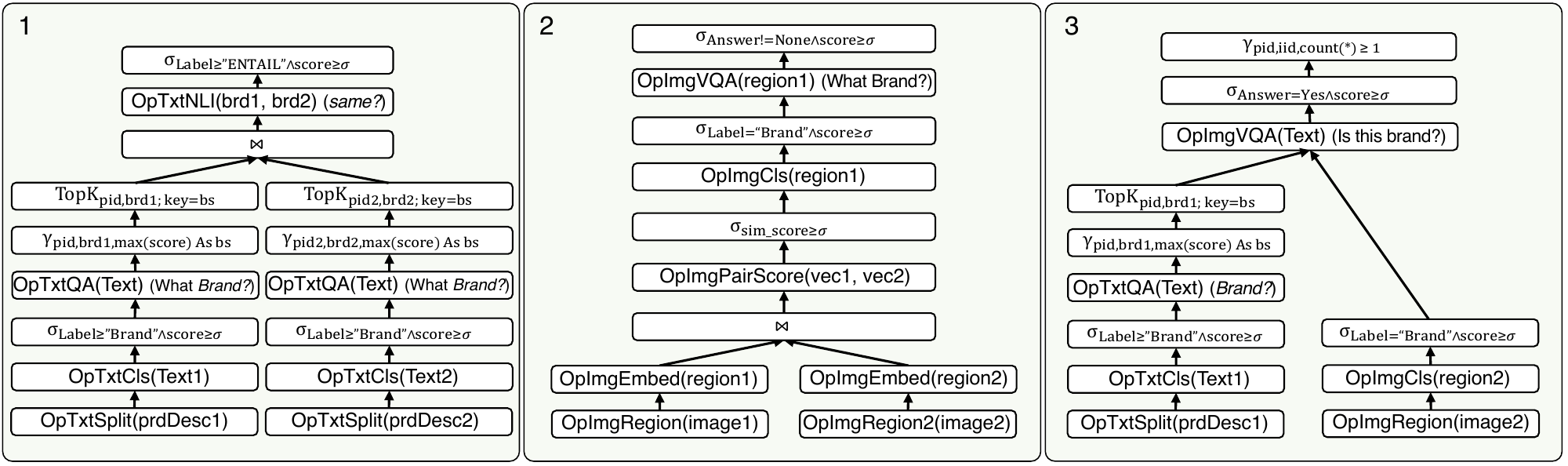}
  \caption{Top-3 logical plans for a representative E-Commerce query.}
  \label{fig:plans}
\end{figure}

\para{Query.} \texttt{SemJoin} query from the E-Commerce scenario:
``You will receive descriptions and images of two products. Determine whether they are from the same brand.
Product A: \{productDisplayName\_right\} \{productDescriptors\_right\}, image \{images\_right\}.
Product B: \{productDisplayName\_left\} \{productDescriptors\_left\}, image \{images\_left\}.''

\para{Interpreting Plans in Figure~\ref{fig:plans}.}
Figure~\ref{fig:plans} shows three logical plans for the representative E-Commerce \texttt{SemJoin} query.
All three plans implement the same intent---deciding whether two products are from the same brand---but they differ in \emph{where} they obtain brand evidence (text vs.\ image) and \emph{how} they make the final match decision.
A key empirical property of this workload is that the brand is often explicitly present in text (e.g., the display name or a few relevant fields inside \texttt{productDescriptors}),
yet \texttt{productDescriptors} is a long, heterogeneous JSON string with many irrelevant keys (e.g., \texttt{materials\_care\_desc}, \texttt{style\_note}, \texttt{inventory}).
Thus, plans that can \emph{isolate brand-bearing fields/segments} from the JSON tend to perform strongly, while plans that rely heavily on image understanding face more noise because product photos frequently contain multiple garments and distracting regions.

\para{Plan~1 (Text-first Brand Matching).}
Plan~1 is a text-only strategy that extracts brand candidates independently from each product's description and then matches them.
Concretely, it (i) applies \textsf{OpTxtSplit} to decompose the long JSON string into manageable segments,
(ii) uses \textsf{OpTxtCls} to keep only segments that likely contain brand cues,
(iii) runs \textsf{OpTxtQA} to extract a brand string with a confidence score from each retained segment,
and (iv) aggregates and retains the top brand candidates per product (Top-$K$).
Finally, it joins the candidate sets across the two products and uses \textsf{OpTxtNLI} to decide whether the two brand strings entail a ``same brand'' relationship.
This plan performs best in our setting because many products already expose the brand in text; once the JSON is decomposed and irrelevant fields are filtered out,
the remaining segments are short and brand-bearing, enabling accurate extraction and a cheap, reliable textual match.
In other words, decomposition primarily addresses the \emph{JSON heterogeneity problem}, after which the join reduces to a near-deterministic comparison.

\para{Plan~2 (Image-first Region Alignment \& Brand Extraction).}
Plan~2 is an image-centric strategy designed for cases where textual cues are missing or unreliable.
It first decomposes each product image into regions via \textsf{OpImgRegion}, embeds regions with \textsf{OpImgEmbed},
and uses \textsf{OpImgPairScore} to align visually similar region pairs across the two products.
It then filters to high-similarity pairs and applies image-side operators (\textsf{OpImgCls} and \textsf{OpImgVQA}) to extract a brand hypothesis from the aligned region(s).
This plan benefits from decomposition because E-Commerce images often include multiple garments or clutter; region-level processing can isolate logo- or label-like patches,
making subsequent inference more focused than operating on the full image.
However, compared to Plan~1 it is typically less consistent on this workload, since brands are frequently conveyed textually and logos may be subtle, partially occluded, or non-textual,
making open-ended brand extraction from images harder and more variable.

\para{Plan~3 (Asymmetric Text-to-Image Verification).}
Plan~3 is a hybrid verification strategy that treats text as the primary source of \emph{candidate} brands and uses images only for \emph{confirmation}.
It first follows the Plan~1 text pipeline to produce a small Top-$K$ set of brand candidates for one (or both) products.
Independently, it decomposes the other product's image into regions and filters to brand-like regions with \textsf{OpImgCls}.
Instead of asking an open-ended ``What brand is this?'' question on the image, it runs \textsf{OpImgVQA} in a \emph{closed-form} manner
(e.g., ``Is this region associated with brand \emph{X}?''), iterating over the text-derived candidates and accepting the pair if any candidate is verified with sufficient confidence.
This plan can be competitive with Plan~1 when text is mostly correct but occasionally ambiguous (e.g., multiple brands mentioned, noisy descriptors),
because decomposition shrinks the text to a concise candidate set and converts image understanding into a simpler yes/no verification task.
Compared to Plan~2, it avoids the brittleness of open-ended brand extraction from images by constraining the answer space to a small set of plausible brands.

\para{Takeaway.}
Across these plans, decomposition is the common enabler: splitting the JSON descriptions into segments and splitting images into regions isolates salient evidence,
reducing both irrelevant context and downstream inference difficulty.
In this workload, because the brand often exists in text, the text-first plan (Plan~1) usually dominates;
image-centric plans are most useful either as alternatives when text cues are absent or as verification mechanisms when text is noisy.

\section{Validation Set Noise} \label{sec:val_noise}

\para{Setup.}
We sanity-check robustness to imperfect validation labels by injecting synthetic noise and rerunning physical planning under the same budget.
We use a binary E-Commerce \texttt{SemJoin} intent that asks whether two product descriptions refer to products of the same category from the same brand (as in our join instruction), and uniformly flip 10\% of validation labels (match $\leftrightarrow$ non-match). We also repeat the experiment with 20\% label flips and observe the same outcome

\begin{table}[ht]
\caption{Top-1 plan selection under synthetic validation-label noise for a binary E-Commerce \texttt{SemJoin} query.}
\vspace{-2mm}
\centering
\small
\setlength{\tabcolsep}{4pt}
\begin{tabular}{l p{10cm} p{5cm}}
\toprule
Validation labels & Top-1 plan (operator sequence) & Differences \\
\midrule
Clean &
Per side:
\texttt{OpTxtSplit(prdDesc)} $\rightarrow$
\texttt{OpTxtCls} $\rightarrow$
\texttt{OpTxtQA(``What Brand?'')} $\rightarrow$
$\gamma$(\texttt{max(score)}) $\rightarrow$
\texttt{TopK} \;\;
Then:
\texttt{OpTxtNLI(brd$_1$,brd$_2$)} $\rightarrow$ $\sigma$(\texttt{ENTAIL} $\wedge$ score$\ge\tau$)
&
Tuned cutpoints only
($\tau_{\mathrm{cls}}, \tau_{\mathrm{qa}}, K, \tau_{\mathrm{nli}}$)
\\[1mm]
10\% flipped &
Same operator sequence (text-first brand extraction $\rightarrow$ textual entailment match)
&
Slightly different cutpoints
($\tau_{\mathrm{cls}}, \tau_{\mathrm{qa}}, K, \tau_{\mathrm{nli}}$)
\\
&
\\
20\% flipped &
Same operator sequence (text-first brand extraction $\rightarrow$ textual entailment match)
&
Slightly different cutpoints
($\tau_{\mathrm{cls}}, \tau_{\mathrm{qa}}, K, \tau_{\mathrm{nli}}$)
\\
\bottomrule
\end{tabular}
\label{tab:val_noise}
\vspace{-3mm}
\end{table}

\para{Result.}
The Top-1 plan remains structurally unchanged under 10\% label flips.
Both runs select the same text-first pipeline that (i) splits the long, heterogeneous \texttt{productDescriptors} into short segments, (ii) filters for brand-bearing segments, (iii) extracts brand candidates with confidence scores, and (iv) matches the two products via an entailment-style comparison of the extracted brand strings.
The only differences are in tuned cutpoints (e.g., confidence thresholds and $K$ in Top-$K$ selection), not in the operator sequence.
Intuitively, alternative candidates (e.g., single-shot classification over the full JSON or overly aggressive early filters) fail systematically on easy cases, so moderate label noise does not change which plan family wins.

\section{\revision{Constrained Optimization Support}} \label{sec:appx:constrained}

\revision{The main results use \OURSYSTEM{}'s default weighted-utility objective. The same optimizer also accepts explicit latency and cost budgets $(L_{\max}, C_{\max})$: each BO trial's sample set metrics are extrapolated to full data units, and a log-scaled penalty on the relative budget violation is added to the scalar utility as $\text{score}(\theta) = U(\theta) - \alpha[\log(1{+}v_L) + \log(1{+}v_C)]$, with $v_L = (L_\theta - L_{\max})_+/L_{\max}$ and $v_C$ defined analogously; $\alpha$ controls the penalty strength. The standard BoTorch acquisition function therefore biases sampling toward budget-respecting regions without any change to the surrogate or acquisition family. We evaluate on the Movie scenario. The \emph{Default} row reproduces the unconstrained result; each constrained row tightens every query's budget by a factor $m$ relative to its observed default $(L_{\text{def}}, C_{\text{def}})$, sweeping $m \in \{0.5, 0.2, 0.1\}$.}

\begin{table}[H]
\caption{\revision{Q, L, and C across queries under \OURSYSTEM{}'s default weighted-utility objective and three constrained variants that tighten each query's budgets to $(L_{\max}, C_{\max}) = m\!\cdot\!(L_{\text{def}}, C_{\text{def}})$. Q is an arithmetic mean across queries; L and C are geometric means.}}
\vspace{-3mm}
\label{tab:appx:constrained}
\centering
\setlength{\tabcolsep}{4pt}
\renewcommand{\arraystretch}{1.05}
\begin{tabular}{lccc}
\toprule
\textbf{Variant} & \textbf{Q}$\uparrow$ & \textbf{L (s)}$\downarrow$ & \textbf{C (\$)}$\downarrow$ \\
\midrule
Default                  & 0.544 & 266 & 0.30 \\
Mild ($m\!=\!0.5$)       & 0.481 & 142 & 0.14 \\
Tight ($m\!=\!0.2$)      & 0.348 &  65 & 0.05 \\
Very tight ($m\!=\!0.1$) & 0.228 &  24 & 0.02  \\
\bottomrule
\end{tabular}
\end{table}

\revision{The logical plan barely changes across $m$; the change is in physical tuning. As budgets tighten, BO progressively reduces the LLM share in favor of local ML and then symbolic backends. At Mild, hard tuples once routed to the LLM fall to local ML, costing modest Q; at 
Tight, the LLM share drops near zero and those tuples fall to symbolic, producing a sharper drop; at Very tight, most tuples are handled 
symbolically, pushing Q toward a floor where easy predicates remain decent but hard joins and extractions collapse. The weighted-utility default is thus a convenience, not a structural 
requirement---users with hard budgets can use the same optimizer by setting $L_{\max}$ and $C_{\max}$.}

\section{\revision{Per-Scenario Quality--Cost Trade-off}} \label{sec:appx:pareto-qc}

\revision{Figures~\ref{fig:pareto-qc-movie}--\ref{fig:pareto-qc-ecomm} report per-scenario Quality--Cost scatter plots for all systems, providing a 2D trade-off view (R3.M1) rather than averaged improvements. Each marker is one (system, query) outcome; up-and-left is better. \OURSYSTEM{} often lies on the upper-left frontier, while the plots also expose
per-query exceptions where a baseline's narrow operator support happens to
match the query.}

\begin{figure}[H]
\centering
\vspace{-2mm}
\includegraphics[width=0.7\linewidth]{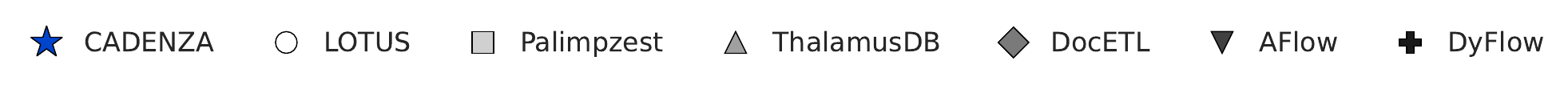}
\vspace{-3mm}
\label{fig:pareto-qc-legend}
\end{figure}

\begin{figure}[H]
\centering
\vspace{-2mm}
\includegraphics[width=0.75\linewidth]{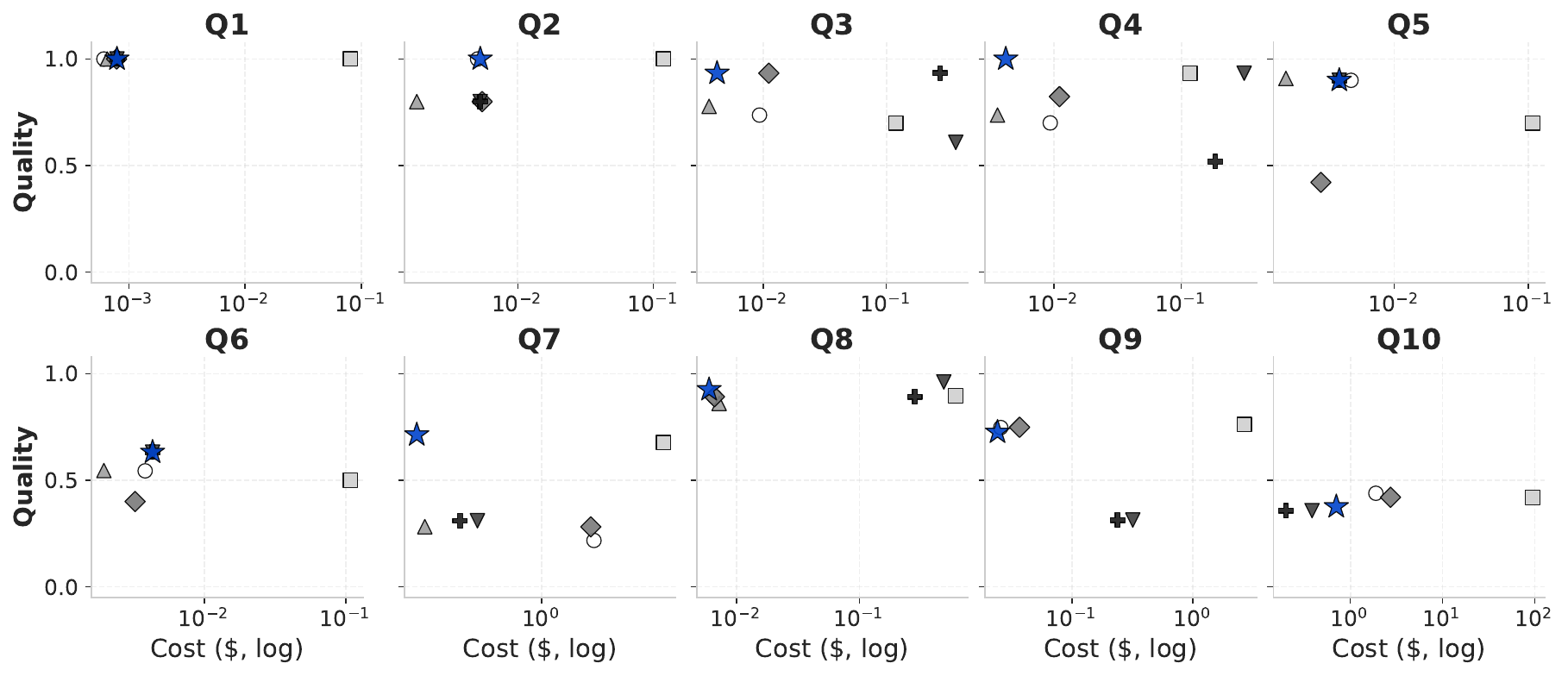}
\vspace{-3mm}
\caption{\revision{Quality--Cost trade-off on Movie. Up-and-left is better.}}
\label{fig:pareto-qc-movie}
\vspace{-3mm}
\end{figure}

\begin{figure}[H]
\centering
\vspace{-2mm}
\includegraphics[width=0.75\linewidth]{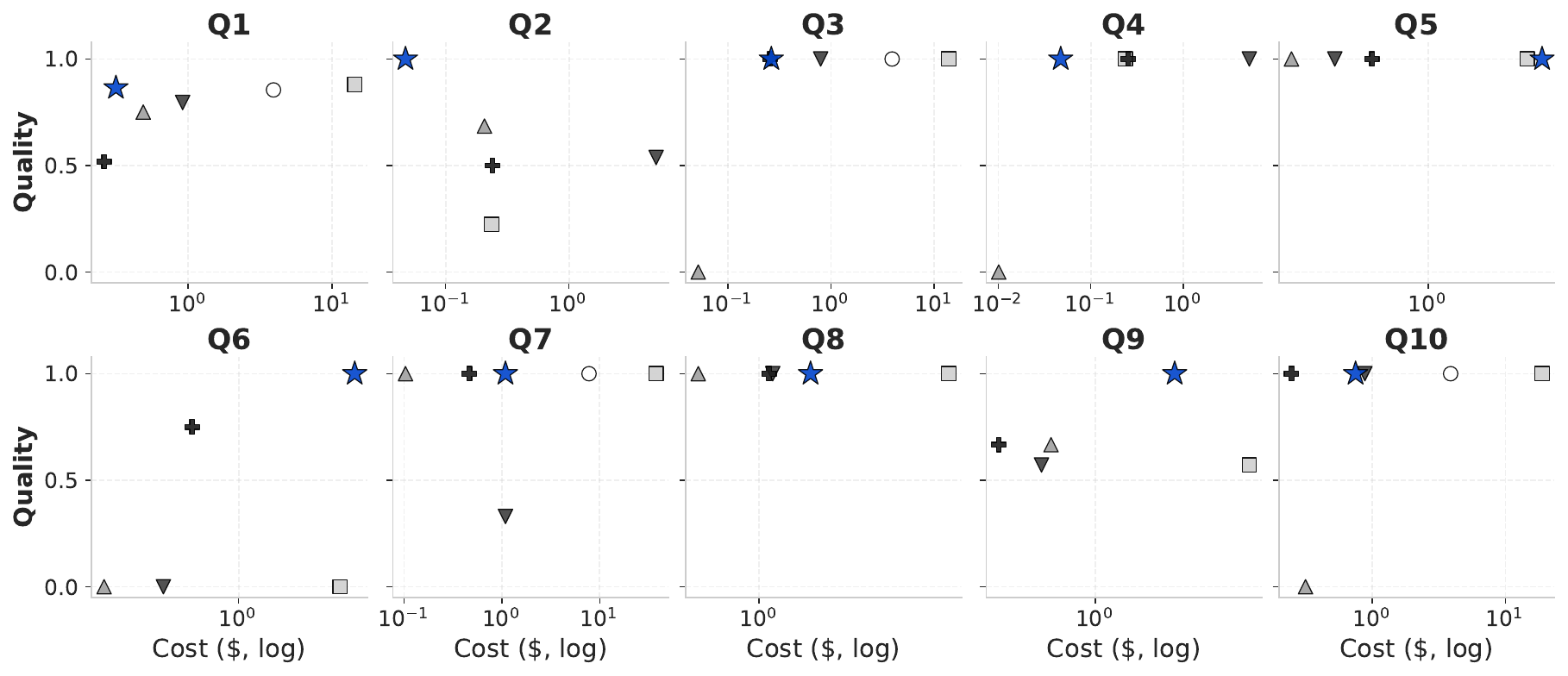}
\vspace{-3mm}
\caption{\revision{Quality--Cost trade-off on Wildlife. Up-and-left is better.}}
\label{fig:pareto-qc-wildlife}
\vspace{-3mm}
\end{figure}

\begin{figure}[H]
\centering
\vspace{-2mm}
\includegraphics[width=0.75\linewidth]{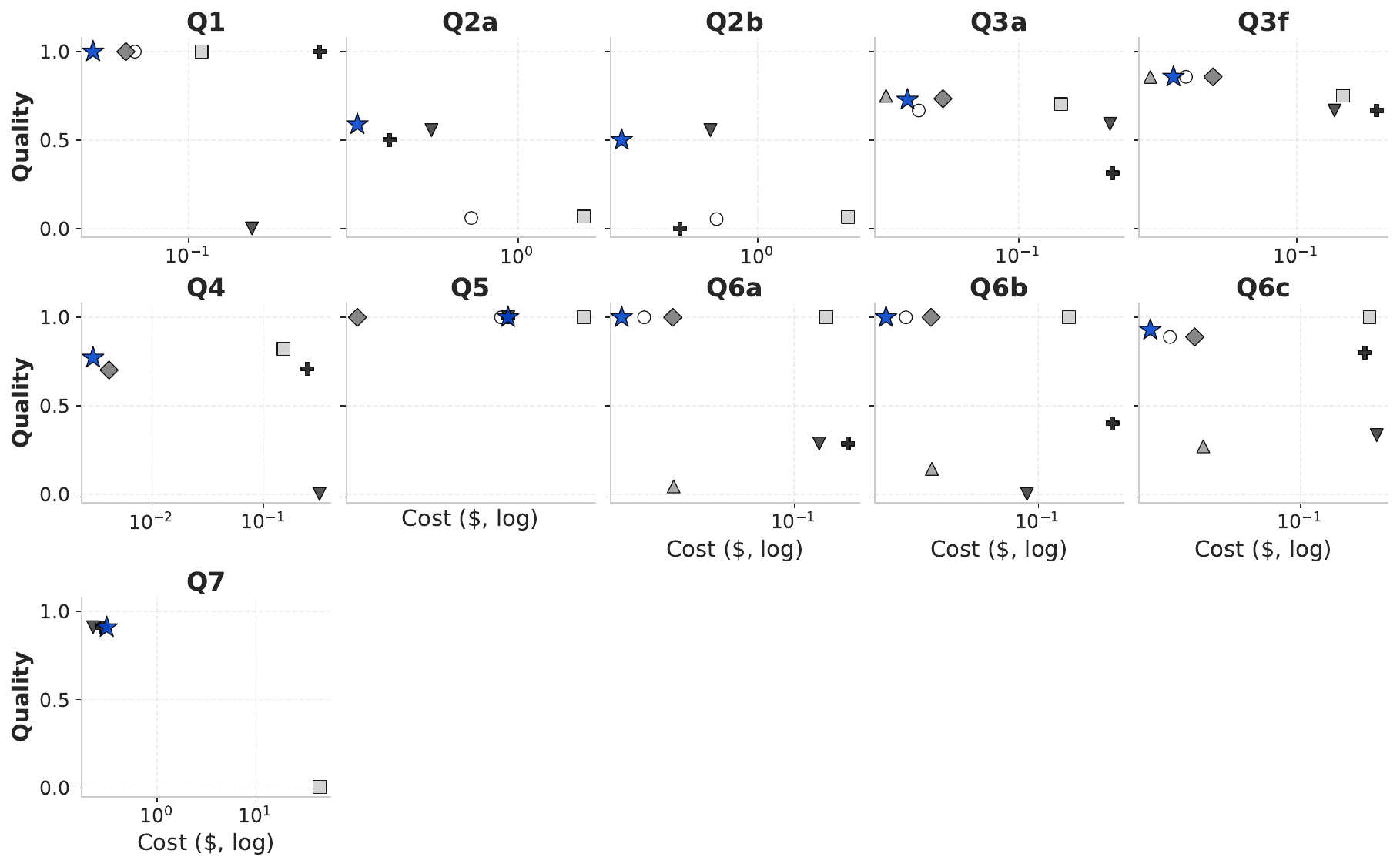}
\vspace{-3mm}
\caption{\revision{Quality--Cost trade-off on MMQA. Up-and-left is better.}}
\label{fig:pareto-qc-mmqa}
\vspace{-3mm}
\end{figure}

\begin{figure}[H]
\centering
\vspace{-2mm}
\includegraphics[width=0.75\linewidth]{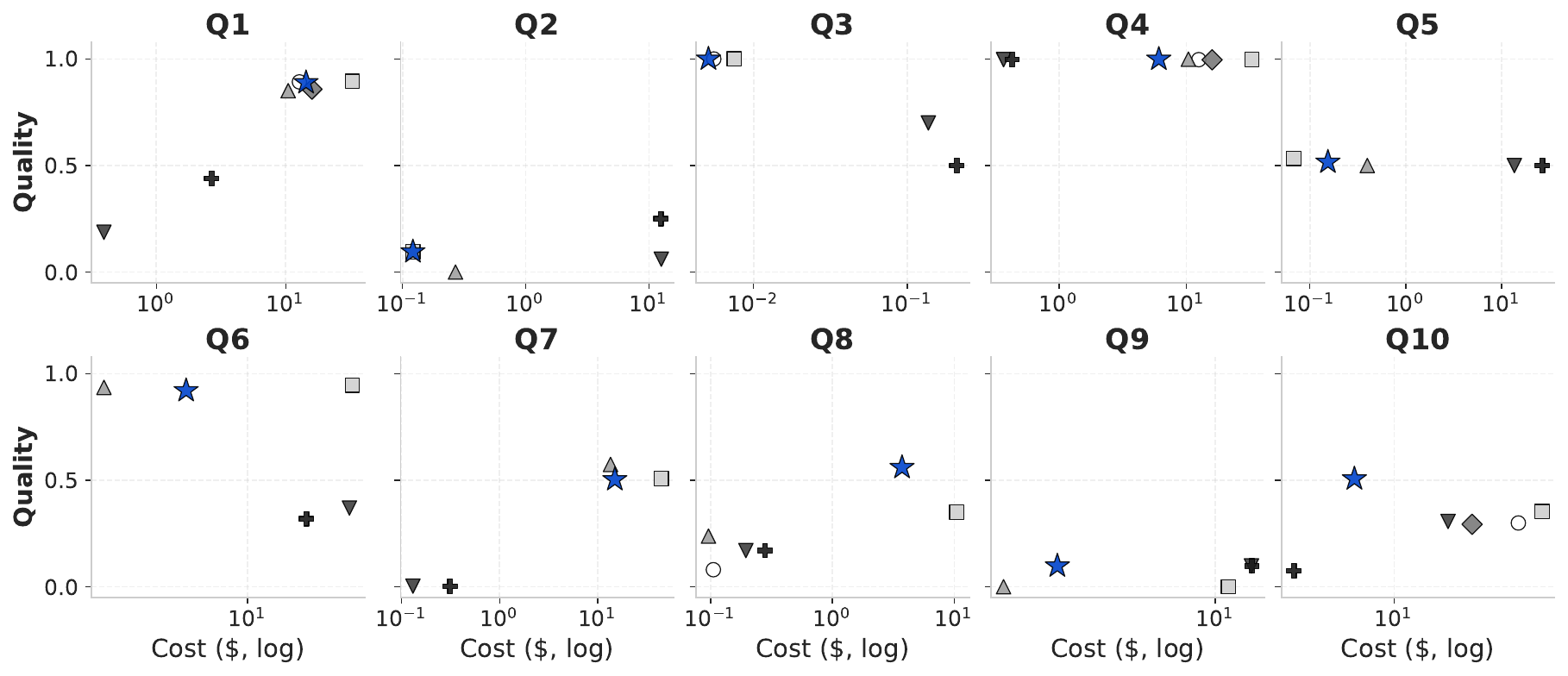}
\vspace{-3mm}
\caption{\revision{Quality--Cost trade-off on Cars. Up-and-left is better.}}
\label{fig:pareto-qc-cars}
\vspace{-3mm}
\end{figure}

\begin{figure}[H]
\centering
\vspace{-2mm}
\includegraphics[width=0.75\linewidth]{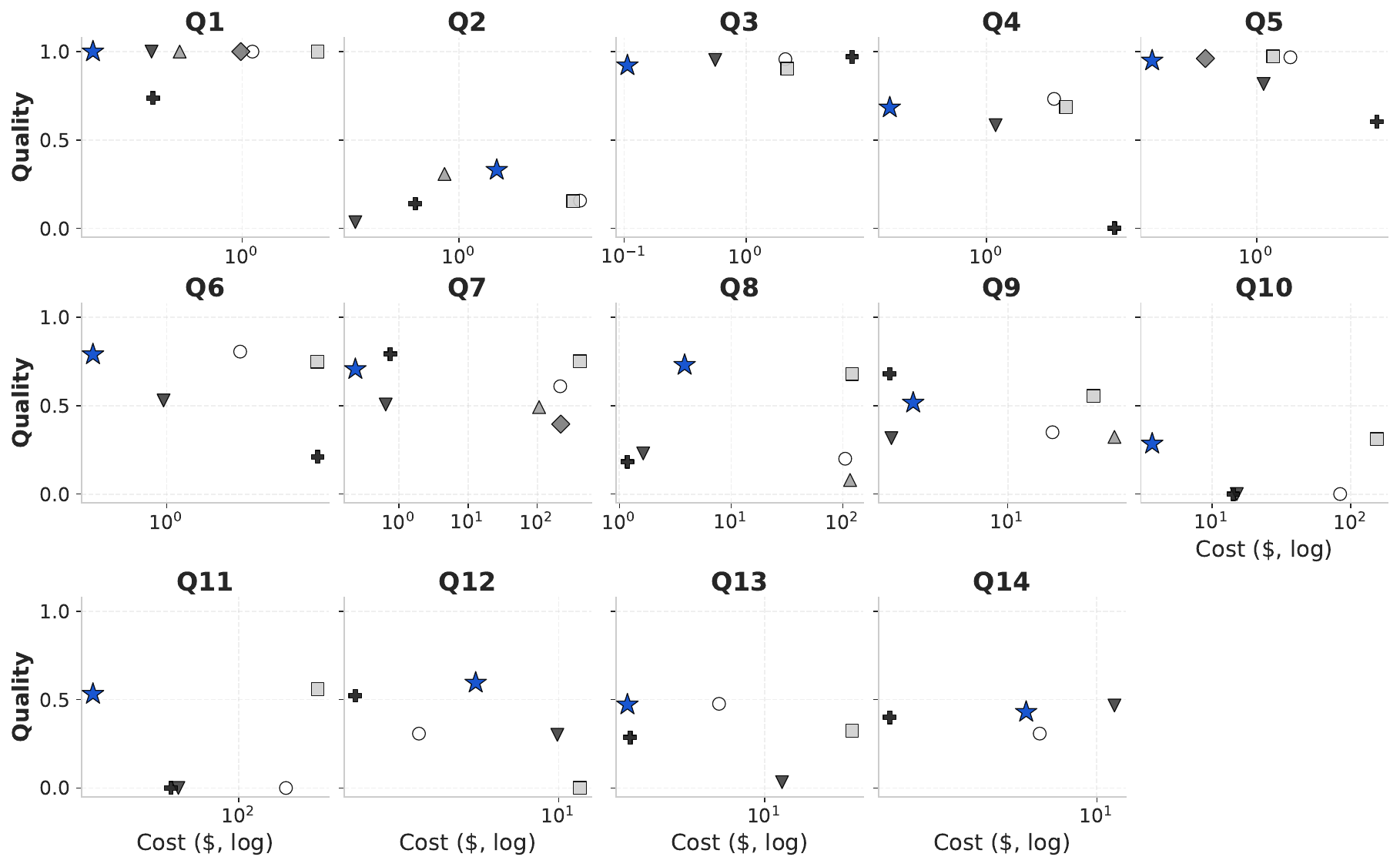}
\vspace{-3mm}
\caption{\revision{Quality--Cost trade-off on E-Commerce. Up-and-left is better.}}
\label{fig:pareto-qc-ecomm}
\vspace{-3mm}
\end{figure}

\clearpage
\section{\revision{Semantic Queries}} \label{sec:appx:queries}

\revision{This section lists the natural-language intent of each semantic query used in our evaluation, together with the core semantic operator chain expressed in LOTUS-style Python where supported, or in \OURSYSTEM{} Python where LOTUS does not natively support the operator. Boilerplate (data loading, path resolution, modality-specific routing knobs) is omitted for brevity; the full source is provided in the accompanying artifact repository.}

\lstdefinestyle{semquery}{
  language=Python,
  basicstyle=\ttfamily\footnotesize,
  breaklines=true,
  breakatwhitespace=true,
  postbreak=\mbox{\textcolor{gray}{$\hookrightarrow$}\space},
  xleftmargin=1.5em,
  frame=none,
  aboveskip=2pt,
  belowskip=4pt,
  showstringspaces=false,
  keywordstyle=\bfseries,
}

\subsection{\revision{Movie}}
\begin{description}[leftmargin=2.5em,labelsep=0.4em,style=nextline,itemsep=2pt]
\item[\textbf{Q1.}] Five clearly positive reviews (any movie). Return \texttt{reviewId}.
\begin{lstlisting}[style=semquery]
Reviews.sem_filter("review is clearly positive: {reviewText}").limit(5)
\end{lstlisting}

\item[\textbf{Q2.}] Five positive reviews for movie \texttt{taken\_3}. Return \texttt{reviewId}.
\begin{lstlisting}[style=semquery]
Reviews[Reviews.id=="taken_3"].sem_filter("review is clearly positive: {reviewText}").limit(5)
\end{lstlisting}

\item[\textbf{Q3.}] Count of positive reviews for movie \texttt{taken\_3}.
\begin{lstlisting}[style=semquery]
len(Reviews[Reviews.id=="taken_3"].sem_filter("review is clearly positive: {reviewText}"))
\end{lstlisting}

\item[\textbf{Q4.}] Positivity ratio of reviews for movie \texttt{taken\_3}.
\begin{lstlisting}[style=semquery]
R = Reviews[Reviews.id=="taken_3"]
ratio = len(R.sem_filter("review is clearly positive: {reviewText}")) / len(R)
\end{lstlisting}

\item[\textbf{Q5.}] Ten pairs of reviews for \texttt{ant\_man\_and\_the\_wasp\_quantumania} expressing the \emph{same} sentiment.
\begin{lstlisting}[style=semquery]
R1.sem_join(R2, "Both reviews share the same sentiment: {R1.reviewText}, {R2.reviewText}",
            on=lambda r1,r2: r1.id==r2.id and r1.reviewId!=r2.reviewId).limit(10)
\end{lstlisting}

\item[\textbf{Q6.}] Ten pairs of reviews for the same movie expressing the \emph{opposite} sentiment.
\begin{lstlisting}[style=semquery]
R1.sem_join(R2, "These two reviews express opposite sentiments: {R1.reviewText}, {R2.reviewText}",
            on=lambda r1,r2: r1.id==r2.id and r1.reviewId!=r2.reviewId).limit(10)
\end{lstlisting}

\item[\textbf{Q7.}] All such opposite-sentiment pairs (no \texttt{LIMIT}).
\begin{lstlisting}[style=semquery]
R1.sem_join(R2, "These two reviews express opposite sentiments: {R1.reviewText}, {R2.reviewText}",
            on=lambda r1,r2: r1.id==r2.id and r1.reviewId!=r2.reviewId)
\end{lstlisting}

\item[\textbf{Q8.}] Counts of positive and negative reviews for \texttt{taken\_3}.
\begin{lstlisting}[style=semquery]
R = Reviews[Reviews.id=="taken_3"].sem_extract({"sentiment": "positive|negative from {reviewText}"})
R.groupby("sentiment").count()
\end{lstlisting}

\item[\textbf{Q9.}] Score 1--5 how much each reviewer liked \texttt{ant\_man\_and\_the\_wasp\_quantumania}.
\begin{lstlisting}[style=semquery]
Reviews[Reviews.id=="ant_man..."].sem_extract({"reviewScore": "int 1-5 from {reviewText}"})
\end{lstlisting}

\item[\textbf{Q10.}] Rank movies by mean per-review score (1--5) over all their reviews.
\begin{lstlisting}[style=semquery]
Reviews.sem_extract({"reviewScore": "int 1-5 from {reviewText}"}).groupby("id").mean()
\end{lstlisting}
\end{description}

\subsection{\revision{Wildlife}}
\begin{description}[leftmargin=2.5em,labelsep=0.4em,style=nextline,itemsep=2pt]
\item[\textbf{Q1.}] How many pictures of zebras do we have?
\begin{lstlisting}[style=semquery]
len(Images.sem_filter("image shows a zebra: {image}"))
\end{lstlisting}

\item[\textbf{Q2.}] How many sound recordings of elephants do we have?
\begin{lstlisting}[style=semquery]
len(Audio.sem_filter("recording is of elephants: {audio}"))
\end{lstlisting}

\item[\textbf{Q3.}] City with the most zebra pictures (ties broken arbitrarily).
\begin{lstlisting}[style=semquery]
Images.sem_filter("image shows a zebra: {image}").groupby("city").count().topk(1)
\end{lstlisting}

\item[\textbf{Q4.}] City with the most elephant audio recordings.
\begin{lstlisting}[style=semquery]
Audio.sem_filter("recording is of elephants: {audio}").groupby("city").count().topk(1)
\end{lstlisting}

\item[\textbf{Q5.}] Cities with elephant images \emph{or} elephant audio.
\begin{lstlisting}[style=semquery]
img = Images.sem_filter("image shows an elephant: {image}").select("city")
aud = Audio.sem_filter("recording is of elephants: {audio}").select("city")
union(img, aud).distinct()
\end{lstlisting}

\item[\textbf{Q6.}] Cities with monkey images but \emph{no} monkey audio.
\begin{lstlisting}[style=semquery]
img = Images.sem_filter("image shows a monkey: {image}").select("city").distinct()
aud = Audio.sem_filter("recording is of monkeys: {audio}").select("city").distinct()
img.except_(aud)
\end{lstlisting}

\item[\textbf{Q7.}] Cities where zebra and impala co-occur in images.
\begin{lstlisting}[style=semquery]
zeb = Images.sem_filter("image shows a zebra: {image}").select("city").distinct()
imp = Images.sem_filter("image shows an impala: {image}").select("city").distinct()
zeb.intersect(imp)
\end{lstlisting}

\item[\textbf{Q8.}] Cities with elephant evidence (image or audio) AND monkey evidence (image or audio).
\begin{lstlisting}[style=semquery]
ele = union(Images.sem_filter("elephant: {image}"), Audio.sem_filter("elephant: {audio}")).select("city")
mon = union(Images.sem_filter("monkey: {image}"), Audio.sem_filter("monkey: {audio}")).select("city")
ele.intersect(mon).distinct()
\end{lstlisting}

\item[\textbf{Q9.}] Cities with both monkey images and monkey audio.
\begin{lstlisting}[style=semquery]
img = Images.sem_filter("image shows a monkey: {image}").select("city").distinct()
aud = Audio.sem_filter("recording is of monkeys: {audio}").select("city").distinct()
img.intersect(aud)
\end{lstlisting}

\item[\textbf{Q10.}] (City, station) with the most zebra pictures.
\begin{lstlisting}[style=semquery]
Images.sem_filter("image shows a zebra: {image}").groupby(["city","station"]).count().topk(1)
\end{lstlisting}
\end{description}

\subsection{\revision{MMQA}}
\begin{description}[leftmargin=2.5em,labelsep=0.4em,style=nextline,itemsep=2pt]
\item[\textbf{Q1.}] Director of the movie that has Ben Piazza as Bob Whitewood.
\begin{lstlisting}[style=semquery]
Movies.sem_join(Roles, "{Movie.synopsis} mentions {Roles.actor} as {Roles.character}")
      .sem_extract({"director": "from {synopsis}"})
\end{lstlisting}

\item[\textbf{Q2a.}] For each racetrack where A.\,P.~Warrior was a contender, find logo images if any.
\begin{lstlisting}[style=semquery]
Racetracks[contender("A.P. Warrior")].sem_join(Images, "{Image} is a logo of {track}")
\end{lstlisting}

\item[\textbf{Q2b.}] Same as Q2a, additionally extract the color of each matched logo.
\begin{lstlisting}[style=semquery]
Q2a_result.sem_extract({"color": "primary color of {Image}"})
\end{lstlisting}

\item[\textbf{Q3a.}] Movies that are comedies.
\begin{lstlisting}[style=semquery]
Movies.sem_filter("movie is a comedy: {synopsis}")
\end{lstlisting}

\item[\textbf{Q3f.}] Movies that are romantic comedies.
\begin{lstlisting}[style=semquery]
Movies.sem_filter("movie is a romantic comedy: {synopsis}")
\end{lstlisting}

\item[\textbf{Q4.}] Categorize each movie by genre (multi-label).
\begin{lstlisting}[style=semquery]
Movies.sem_extract({"genres": "list of genres from {synopsis}"})
      .explode("genres")
\end{lstlisting}

\item[\textbf{Q5.}] Actor common to a fixed list of 16 movies.
\begin{lstlisting}[style=semquery]
Cast[Cast.movie.isin(MOVIE_LIST)].groupby("actor").count() == 16
\end{lstlisting}

\item[\textbf{Q6a.}] Airlines with destinations in Frankfurt.
\begin{lstlisting}[style=semquery]
Airlines.sem_filter("airline serves Frankfurt: {destinations}")
\end{lstlisting}

\item[\textbf{Q6b.}] Airlines with destinations in Germany.
\begin{lstlisting}[style=semquery]
Airlines.sem_filter("airline serves any destination in Germany: {destinations}")
\end{lstlisting}

\item[\textbf{Q6c.}] Airlines with destinations in Europe.
\begin{lstlisting}[style=semquery]
Airlines.sem_filter("airline serves any destination in Europe: {destinations}")
\end{lstlisting}

\item[\textbf{Q7.}] For each airline with European destinations, find its logo (if any).
\begin{lstlisting}[style=semquery]
Q6c_result.sem_join(Images, "{Image} is the logo of {Airline.name}")
\end{lstlisting}
\end{description}

\subsection{\revision{Cars}}
\begin{description}[leftmargin=2.5em,labelsep=0.4em,style=nextline,itemsep=2pt]
\item[\textbf{Q1.}] Cars whose textual complaint entails a crash/accident/collision.
\begin{lstlisting}[style=semquery]
Complaints.sem_filter("complaint entails the car was in a crash/accident/collision: {summary}")
\end{lstlisting}

\item[\textbf{Q2.}] Cars whose audio recording shows a dead battery.
\begin{lstlisting}[style=semquery]
(Cars |><| Audio).sem_filter("audio recording: car has a dead battery: {audio_path}", oracle="audio")
\end{lstlisting}

\item[\textbf{Q3.}] Ten cars with Manual transmission whose image shows the car is \emph{not} damaged.
\begin{lstlisting}[style=semquery]
((Cars |><| Images)[transmission=="Manual"]
  .sem_filter("image: car is not damaged: {image_path}")).limit(10)
\end{lstlisting}

\item[\textbf{Q4.}] Average age of cars whose complaint mentions engine problems.
\begin{lstlisting}[style=semquery]
mean(2026 - (Cars |><| Complaints)
       .sem_filter("complaint: car has engine problem: {summary}").year)
\end{lstlisting}

\item[\textbf{Q5.}] Cars damaged based on \emph{both} audio and image evidence.
\begin{lstlisting}[style=semquery]
aud = Audio.sem_filter("audio captures damaged car: {audio_path}", oracle="audio")
img = Images.sem_filter("image: car is damaged: {image_path}")
aud.merge(img, on="car_id")
\end{lstlisting}

\item[\textbf{Q6.}] Cars with damage evidence in at least two modalities (image, audio, text).
\begin{lstlisting}[style=semquery]
img = Images.sem_filter("image: car is damaged: {image_path}")
aud = Audio.sem_filter("audio captures damaged car: {audio_path}", oracle="audio")
txt = Complaints.sem_filter("complaint: car was on fire or burned: {summary}")
multimodal_at_least_two(img, aud, txt)
\end{lstlisting}

\item[\textbf{Q7.}] Cars that are dented (image), have worn brakes (audio), \emph{and} report electrical issues (text).
\begin{lstlisting}[style=semquery]
dented = (Cars |><| Images).sem_filter("image: car is dented: {image_path}")
worn   = (Cars |><| Audio ).sem_filter("audio: car has worn brakes: {audio_path}", oracle="audio")
elec   = (Cars |><| Complaints).sem_filter("complaint: electrical-system problem: {summary}")
intersect(dented, worn, elec)
\end{lstlisting}

\item[\textbf{Q8.}] First 100 cars whose image shows both punctures and paint scratches.
\begin{lstlisting}[style=semquery]
(Cars |><| Images).sem_filter("image: car has both puncture and paint scratches: {image_path}").head(100)
\end{lstlisting}

\item[\textbf{Q9.}] Cars whose image shows tearing \emph{and} whose audio shows bad ignition.
\begin{lstlisting}[style=semquery]
torn = Images.sem_filter("image: car is torn: {image_path}")
ign  = Audio.sem_filter("audio: car has bad ignition: {audio_path}", oracle="audio")
torn.merge(ign, on="car_id")
\end{lstlisting}

\item[\textbf{Q10.}] Classify each complaint into a problem category.
\begin{lstlisting}[style=semquery]
Complaints.sem_extract({"problem_category": "from {summary}"})
\end{lstlisting}
\end{description}

\subsection{\revision{E-Commerce}}
\begin{description}[leftmargin=2.5em,labelsep=0.4em,style=nextline,itemsep=2pt]
\item[\textbf{Q1.}] Product ids of backpacks from Reebok (text).
\begin{lstlisting}[style=semquery]
Products.sem_filter("product is a backpack from Reebok: {description}")
\end{lstlisting}

\item[\textbf{Q2.}] Sports shoes that are predominantly yellow and silver (image).
\begin{lstlisting}[style=semquery]
Products.sem_filter("image shows sports shoes that are predominantly yellow and silver: {image}")
\end{lstlisting}

\item[\textbf{Q3.}] For each product, extract the brand name from text.
\begin{lstlisting}[style=semquery]
Products.sem_extract({"category": "brand name from {title}, {description}"})
\end{lstlisting}

\item[\textbf{Q4.}] For each product, extract the primary color from the image.
\begin{lstlisting}[style=semquery]
Products.sem_extract({"category": "primary color of {image}"})
\end{lstlisting}

\item[\textbf{Q5.}] Classify each product into one of \{Dress, Bottomwear, Socks, Topwear, Innerwear\} from text.
\begin{lstlisting}[style=semquery]
Products.sem_extract({"category": "one of {Dress|Bottomwear|Socks|Topwear|Innerwear} from {title}, {description}"})
\end{lstlisting}

\item[\textbf{Q6.}] Same five-class classification, from the product image.
\begin{lstlisting}[style=semquery]
Products.sem_extract({"category": "one of {Dress|Bottomwear|Socks|Topwear|Innerwear} from {image}"})
\end{lstlisting}

\item[\textbf{Q7.}] Self-join of products with price~$\le 500$ on same article-type and brand (text-to-text).
\begin{lstlisting}[style=semquery]
P = Products[price <= 500]
P.sem_join(P, "two products are of the same article type from the same brand: {P1.description}, {P2.description}",
           on=lambda p1,p2: p1.id != p2.id)
\end{lstlisting}

\item[\textbf{Q8.}] Self-join: for each product description $\ge$ 3000 chars, find matching product images.
\begin{lstlisting}[style=semquery]
Products[len(description) >= 3000]
  .sem_join(Products, "image {Image} matches the description {description}, title {title}")
\end{lstlisting}

\item[\textbf{Q9.}] Image-to-image self-join finding similar products by base color and article type.
\begin{lstlisting}[style=semquery]
Products.sem_join(Products, "two product images are of the same article type and base color: {image1}, {image2}",
                  on=lambda p1,p2: p1.id != p2.id)
\end{lstlisting}

\item[\textbf{Q10.}] 3-way join: shoes $\bowtie$ bottomwear $\bowtie$ topwear matching by color and brand (cross-modal).
\begin{lstlisting}[style=semquery]
shoes = Products[isFootwear].sem_filter("...")
lower = Products[isBottomwear].sem_filter("...")
upper = Products[isTopwear].sem_filter("...")
sem_join_chain(shoes, lower, upper, on="color,brand")
\end{lstlisting}

\item[\textbf{Q11.}] 4-way join across multiple categories on color and brand (cross-modal).
\begin{lstlisting}[style=semquery]
sem_join_chain(shoes, bottomwear, topwear, accessories, on="color,brand")
\end{lstlisting}

\item[\textbf{Q12.}] For each item, generate JSON \texttt{\{id, brand, category\}} via combined map+classify on text \emph{and} image.
\begin{lstlisting}[style=semquery]
Products.sem_extract({"brand": "...", "category": "topwear|bottomwear from {title},{description},{image}"})
        .map(lambda r: {"id": r.id, "brand": r.brand, "category": r.category})
\end{lstlisting}

\item[\textbf{Q13.}] Multi-modal complex filter (e.g., men's sports tshirts; blue/black; short sleeves; round neck; striped; not winter).
\begin{lstlisting}[style=semquery]
Products.sem_filter("matches all attributes: men, sports, tshirt, blue or black, short-sleeve, round-neck, striped, not winter season: {description}, {image}")
\end{lstlisting}

\item[\textbf{Q14.}] For each product priced $<130$, find the single best-matching image, but only if it depicts white socks.
\begin{lstlisting}[style=semquery]
Products[price < 130].sem_topk(Images, "image best matches the product and depicts white socks: {description}, {Image}", k=1)
\end{lstlisting}
\end{description}

\subsection{\revision{BioDEX}}

\revision{For BioDEX, we follow LOTUS~\cite{LOTUS}'s formulation, which 
casts the task---extreme multi-label classification of 
adverse drug reactions reported in medical articles---as a single 
\texttt{sem\_join} between articles and reactions. Unlike the SemBench 
scenarios, which contain multiple distinct queries, BioDEX therefore 
appears here as one query.}

\begin{description}[leftmargin=2.5em,labelsep=0.4em,style=nextline,itemsep=2pt]
\item[\textbf{Q.}] For each medical article, identify the adverse drug reactions reported, modeled as an article\,$\bowtie$\,reaction \texttt{sem\_join}.
\begin{lstlisting}[style=semquery]
articles.sem_join(reactions,
                  "the article {article} reports the adverse drug reaction {reaction}")
\end{lstlisting}
\end{description}

\fi

\end{document}